\documentclass[aps,prb,twocolumn]{revtex4}

\newcommand{\bk}{{\bf k}}

\newcommand{\be}{\begin{equation}}
\newcommand{\ee}{\end{equation}}
\newcommand{\bea}{\begin{eqnarray}}
\newcommand{\eea}{\end{eqnarray}}

\newcommand{\lMo}{\lambda_{\rm Mo}}
\newcommand{\lS}{\lambda_{\rm S}}

\newcommand{\bwt}{\begin{widetext}}
\newcommand{\ewt}{\end{widetext}}

\usepackage{graphics,graphicx,amsmath}
\usepackage{tabularx,float}
\usepackage{epsfig}

\usepackage{bm,bbold}
\usepackage{color}
\usepackage{verbatim}

\usepackage[T1]{fontenc}
\usepackage{libertine}

\begin{document}

\title{Momentum dependence of spin-orbit interaction effects in single-layer and multi-layer transition metal dichalcogenides}

\author{R. Rold\'an, M.P. L\'opez-Sancho, F. Guinea}
\address{Instituto de Ciencia de Materiales de Madrid,
CSIC, c/ Sor Juana Ines de la Cruz 3, 28049 Cantoblanco, Madrid, Spain}

\author{E. Cappelluti}
\address{Istituto de Sistemi Complessi, U.O.S. Sapienza, CNR, v. dei
Taurini 19, 00185 Roma, Italy}

\author{J.A. Silva-Guill\'en, P. Ordej\'on}
\address{ICN2 - Institut Catala de Nanociencia i Nanotecnologia, Campus UAB, 08193 Bellaterra, Spain}
\address{CSIC - Consejo Superior de Investigaciones Ciaentificas, ICN2 Building, 08193 Bellaterra, Spain}

\begin{abstract}
One of the main characteristics of the new family of two-dimensional crystals of semiconducting transition metal dichalcogenides (TMD) is the strong spin-orbit interaction, which makes them very promising for future applications in spintronics and valleytronics devices. 
Here we present a detailed study of the effect of spin-orbit coupling (SOC)
on the band structure of single-layer and bulk TMDs, including explicitly the role of the
chalcogen orbitals and their hybridization with the transition metal atoms.
To this aim, we combine density functional theory (DFT) calculations with a Slater-Koster tight-binding model. Whereas most of the previous tight-binding models have been restricted to the K and K' points of the Brillouin zone (BZ), here we consider the effect of SOC in the whole BZ, and the results are compared to the band structure obtained by DFT methods. The tight-binding model is used to analyze the effect of SOC in the band structure, considering separately the contributions from the transition metal and the chalcogen atoms.  Finally, we present a scenario where, in the case of strong
SOC, the spin/orbital/valley entanglement at
the minimum of the conduction band at Q can be probed
and be of experimental interest in 
the most common cases of electron-doping reported for this family of compounds.   
\end{abstract}

\date{\today}
\maketitle

\section{Introduction}

Transition metal dichalcogenides have emerged as a new family of layered materials with a number of remarkable electrical and optical properties.\cite{WS12} Among them, single layers of the semiconducting compounds of the group-VIB {\em MX}$_2$ (where $M=$ Mo, W and $X=$ S, Se) are of special interest because they have a direct band gap in the visible range of the spectrum,\cite{MH10} which is located in the K and K' points of the hexagonal BZ.\cite{CG13} The absence of inversion symmetry in single layer samples lifts the spin degeneracy of the energy bands in the presence of SOC.\cite{ZCS11} Interestingly, the spin splitting in inequivalent valleys must be opposite, as imposed by time reversal symmetry. This leads to the so called spin-valley coupling,\cite{XY12} which has been studied theoretically\cite{FX12,SX13,RMA13,RGP13,LX13} and observed experimentally.\cite{Cao_etal_2012,Zeng_etal_2012,Mak_etal_2012,WX13,WS13,ZC13} Although the SOC splitting of the bands is particularly large in the valence band ($\sim 150$ meV for MoS$_2$ and $\sim 400$ meV for WS$_2$), a finite SOC splitting of the conduction band is also allowed by symmetry,\cite{OR13} as confirmed by recent density functional theory  calculations.\cite{CL12,Kosmider_etal_2013,MW13,SD13,KF13,KB13,KGF13} 
In addition, interlayer coupling plays here also a fundamental role.
Indeed, the band structure dramatically changes from
single-layer to multi-layer samples, involving a transition from a
direct gap for single-layer samples to an indirect gap for multi-layer
samples,\cite{CG13} as it has been observed experimentally.\cite{MH10,ZE13,JO13,ZS14} 

Both numerical first-principles techniques and analytical
  approaches have been employed to investigate the role of the SOC in
  these materials. Within this context, the SOC has
  been mainly included in tight-binding (TB) models valid only in the low-energy
  range, where the presence of the $p$-orbitals of the chalcogen
  atoms has been integrated out in an effective model (Refs. \onlinecite{XY12,LZN13,KL13,WW14,COG13,GMP14}).
Alternatively, DFT calculations can provide  a more compelling description,
but their complexity hampers the extraction of a simple model of the SOC.
From a more general point of view, finally,
most of the recent works on the effects of SOC in TMDs have been focused
on single-layer samples,
whereas fewer investigations have been devoted to the effect of
SOC on the band structure of multi-layer and bulk samples.
In particular, a complete TB model that can account for the
effect of SOC in the whole BZ,
including explicitly the $p$-orbitals of the chalcogen atoms,
 is lacking. Such a TB model is especially useful to study cases
 where  DFT methods result too challenging computationally, as the
effect of disorder, inhomogeneous strain, strong many-body interactions, etc.

In this paper we use
a combination of TB and DFT calculations to provide
a complete TB model, in the whole BZ, of the effects of SOC on the band structure of
single-layer and multi-layer TMD
taking explicitly into account the $p$-orbitals of the chalcogen atoms,
and the atomic spin-orbit interaction on both the metal and chalcogen orbitals. 
The bands obtained from the TB model are compared
to the corresponding DFT band structure for single layer
and bulk MoS$_2$ and WS$_2$. By considering the main orbital contribution at each relevant point of the BZ, we analyze the origin and main features of the SOC effects at the different band edges.
Such model provides a useful base not only for the analytical investigation
of the role of the SOC in the presence of local strain tuning
the $M$-$X$ distance, but also for the investigation of the
microscopical relevant spin-orbit processes. In particular,
we show that the terms associated to second order
spin-flip processes of the SOC can be safely neglected for most of the cases of experimental interest. The TB model developed here is especially  useful to analyze the effect of SOC at the so-called Q point of the BZ, which corresponds to the absolute minimum of the conduction band of multi-layer samples.
We finally discuss also the peculiarities of the SOC in bilayer
{\em MX}$_2$, for which the spin-valley-layer coupling could be exploit for future valleytronics applications.

The paper is organized as follows.
In Section II we present the  model  for the single layer and bulk cases.
The  comparison between TB and DFT band structures
considering the SOC effects, is illustrated in Section III for
MoS$_2$ and WS$_2$.  Results are presented and discussed
in Section IV.  Finally the main findings are summarized
 and some  conclusions are given in Section V.

\section{Spin-orbit interaction and the tight-binding Hamiltonian}
\label{s:model}

In this section we present the analytical structure
of the TB Hamiltonians for single-layer and bulk TMD {\em MX}$_2$
compounds including the SO interaction.
Specific parameters for realistic materials will be provided
in the next section, as well as a discussion of the physical
consequences of the SOC.

\subsection{Single-layer case}\label{Sec:SL}

The TMD $MX_2$ are composed, in their bulk configuration, of two-dimensional $X-M-X$ layers stacked on top of each other, coupled by weak van der Waals forces. The $M$ atoms are ordered in a triangular lattice, each of them bonded to six $X$ atoms located in the top and bottom layers, forming a sandwiched material. Our starting point will be a 11-band TB spinless model which, for the single-layer, considers the
five $d$ orbitals of the metal atom $M$ and the three $p$ orbitals
for each of the two chalcogen atoms $X$ in the top and bottom layer.\cite{CG13}
We can introduce a Hilbert base defined by the 11-fold vector:
\begin{eqnarray}
\phi_i^\dagger
&=&
(
p_{i,x,t}^\dagger,
p_{i,y,t}^\dagger,
p_{i,z,t}^\dagger,
d_{i,3z^2-r^2}^\dagger,
d_{i,x^2-y^2}^\dagger,\nonumber\\
&&
d_{i,xy}^\dagger,
d_{i,xz}^\dagger,
d_{i,yz}^\dagger,
p_{i,x,b}^\dagger,
p_{i,y,b}^\dagger,
p_{i,z,b}^\dagger
),
\label{basis}
\end{eqnarray}
where $d^{\dagger}_{i,\alpha}$ creates an electron 
in the orbital $\alpha$ of the $M$ atom in the $i$-unit cell,
$p^{\dagger}_{i,\alpha,t}$ creates an electron 
in the orbital $\alpha$ of the top ($t$) layer atom $X$ in the $i$-unit cell,
and $p^{\dagger}_{i,\alpha,b}$ creates an electron 
in the orbital $\alpha$ of the bottom ($b$) layer atom $X$ in the $i$-unit
cell. After an appropriate unitary transformation,
the spinless (sl) representation of the 
single-layer (1L) Hamiltonian can be expressed in the block form
\begin{eqnarray}
\hat{H}^{\rm sl}_{\rm 1L}( {\bf k})
&=&
\left(
\begin{array}{cc}
\hat{H}_{\rm E} &  \hat{0}_{6\times 5} \\
\hat{0}_{5\times 6} & \hat{H}_{\rm O} 
\end{array}
\right),
\label{Eq:H0}
\end{eqnarray}
where $\hat{H}_{\rm E} $ and $\hat{H}_{\rm O} $ are a $6\times 6$ and
$5\times 5$ blocks with even (E) and odd (O) parity respectively
upon the mirror inversion $z\rightarrow -z$, and $\hat{0}_{m\times n}$ denotes $m\times n$ zero matrices.\cite{CG13} In
particular, $\hat{H}_{\rm E}$ is built from hybridizations of the $d_{xy}$,
$d_{x^2-y^2}$, $d_{3z^2-r^2}$ orbitals of the metal $M$ with
the symmetric (antisymmetric) combinations of the $p_x$, $p_y$ ($p_z$)
orbitals of the top and bottom chalcogen atoms $X$. On the other hand,
the {\it odd} block, $\hat{H}_{\rm O}$, is made by hybridizations of the
$d_{xz}$ and $d_{yz}$
orbitals of $M$ with the antisymmetric (symmetric) combinations
of the $p_x$, $p_y$ ($p_z$) orbitals of the $X$ atom
in the top and bottom layers.
Explicit expressions for all the matrix elements 
in terms of the Slater-Koster parameters were obtained in
Ref. \onlinecite{CG13}, and we notice that the $6\times6$ {\it even} block
$\hat{H}_{\rm E}$ contains the relevant orbital contribution for the states of the upper valence band and the lower conduction band.

In the context of the present TB model, we include the
SOC term in the Hamiltonian by means
of a pure atomic spin-orbit interaction acting on both the metal 
and chalcogen atoms.
Explicitly we consider here the SOC given by:
\be
\hat{H}^{\rm SO}=\sum_a \frac{\lambda_a}{\hbar}
{\bf \hat{L}}_a\cdot {\bf \hat{S}}_a,
\ee 
where $\lambda_a$, the intra-atomic SOC constant, depends on the specific atom ($a=M,X$).
${\bf \hat{L}}_a$ is the atomic orbital angular momentum operator
and
${\bf \hat{S}}_a$ is the 
electronic spin operator.\cite{GM99,CLM04,HGB06} It is convenient to use the representation 
\be\label{Eq:HSO}
\hat{H}^{\rm SO}
=
\sum_a \frac{\lambda_a}{\hbar} \left(
\frac{\hat{L}_a^{+}\hat{S}_a^{-}
+\hat{L}_a^{-}\hat{S}_a^{+}}{2}+\hat{L}_a^z\hat{S}_a^z 
\right),
\ee
where (omitting now for simplicity the atomic index $a$):
\begin{equation}
\hat{S}^+
=
\left(
\begin{array}{cc}
0 &  1 \\
0 & 0
\end{array}
\right), 
\quad
\hat{S}^-
=
\left(
\begin{array}{cc}
0 &  0 \\
1 & 0
\end{array}
\right),
\quad
\hat{S}^z
=
\frac{1}{2}\left(
\begin{array}{cc}
1 &  0 \\
0 & -1
\end{array}
\right).
\label{Eq:SMatrices}
\end{equation}
In a similar way,
the orbital angular momentum operator ${\bf \hat{L}}$
acts on the states $|l,m\rangle$
as
\bea
\hat{L}^{\pm}|l,m\rangle& = &\hbar\sqrt{l(l+1)-m(m\pm 1)}\;|l,m\pm 1\rangle,
\nonumber\\
\hat{L}^z| l,m\rangle & = & \hbar m \;|l,m\rangle,
\eea
where $l$ refers to the orbital momentum quantum number and
$m$ to its $z$ component.

We choose the orbital basis set in the following manner:
\bea
|p_z\rangle&=&|1,0\rangle\nonumber\\
|p_x\rangle&=&-\frac{1}{\sqrt{2}}[|1,1\rangle-|1,-1\rangle]\nonumber\\
|p_y\rangle&=&\frac{i}{\sqrt{2}}[|1,1\rangle+|1,-1\rangle]\nonumber\\
|d_{3z^2-r^2}\rangle&=&|2,0\rangle\nonumber\\
|d_{xz}\rangle&=&-\frac{1}{\sqrt{2}}[|2,1\rangle-|2,-1\rangle]\nonumber\\
|d_{yz}\rangle&=&\frac{i}{\sqrt{2}}[|2,1\rangle+|2,-1\rangle]\nonumber\\
|d_{x^2-y^2}\rangle&=&\frac{1}{\sqrt{2}}[|2,2\rangle+|2,-2\rangle]\nonumber\\
|d_{xy}\rangle&=&-\frac{i}{\sqrt{2}}[|2,2\rangle-|2,-2\rangle]
\eea
We further simplify the problem by introducing the aforementioned symmetric (S) and antisymmetric (A) combination of the $p$ orbitals of the top ($t$) and bottom ($b$) $X$ layers:
\bea
|p_{\alpha,S}\rangle&=&\frac{1}{\sqrt{2}}[|p_{\alpha,t}\rangle+|p_{\alpha,b}\rangle],\nonumber\\
|p_{\alpha,A}\rangle&=&\frac{1}{\sqrt{2}}[|p_{\alpha,t}\rangle-|p_{\alpha,b}\rangle].
\eea

The total Hamiltonian, including the SO interaction for the single-layer, can be now written as
\be\label{Eq:Htotal}
\hat{H}_{\rm 1L}({\bf k})
=
\hat{H}^{\rm sl}_{\rm 1L}({\bf k})\otimes {\mathbb 1}_2
+ \hat{H}^{\rm SO}_{\rm 1L},
\ee
where the SOC term $\hat{H}^{\rm SO}_{\rm 1L}$ is
\begin{eqnarray}
\hat{H}^{\rm SO}_{\rm 1L}
&=&
\left(
\begin{array}{cc}
\hat{M}^{\uparrow\uparrow}  &  \hat{M}^{\uparrow\downarrow}\\
\hat{M}^{\downarrow\uparrow} & \hat{M}^{\downarrow\downarrow} 
\end{array}
\right),
\label{Eq:MatrixSO}
\end{eqnarray}
and where
\begin{eqnarray}
\hat{M}^{\sigma\sigma}
&=&
\left(
\begin{array}{cc}
\hat{M}_{\rm EE}^{\sigma\sigma} &  \hat{0}_{6\times5} \\
\hat{0}_{5\times6} &  \hat{M}_{\rm OO}^{\sigma\sigma}
\end{array}
\right),
\label{Eq:MatrixSO_diag}
\end{eqnarray}
and
\begin{eqnarray}
\hat{M}^{\sigma\bar{\sigma}}
&=&
\left(
\begin{array}{cc}
\hat{0}_{6\times 6} &  \hat{M}_{\rm EO}^{\sigma\bar{\sigma}}\\
\hat{M}_{\rm OE}^{\sigma\bar{\sigma}} & \hat{0}_{5\times5}  \\
\end{array}
\right).
\label{Eq:MatrixSO_nodiag}
\end{eqnarray}
Here we have chosen the spin notation $\bar{\sigma}=\downarrow$
($\bar{\sigma}=\uparrow$) when $\sigma=\uparrow$ ( $\sigma=\downarrow$).

The different blocks $\hat{M}_{\rm EE}^{\sigma\sigma}$,
$\hat{M}_{\rm OO}^{\sigma\sigma}$,
$\hat{M}_{\rm EO}^{\sigma\bar{\sigma}}$,
$\hat{M}_{\rm OE}^{\sigma\bar{\sigma}}$,
that constitute the above $22\times22$
matrix, are explicitly reported in the Appendix \ref{App:SOC}.
We notice here that,
in the most general case, the SO interaction couples the E and O
sectors of the
$22 \times 22$ TB matrix. Such mixing 
arises in particular from the spin-flip/spin-orbital processes
associated with the transverse quantum fluctuation described
 by the first two terms of Eq. (\ref{Eq:HSO}).
The effective relevance of these terms can now be directly
investigated in a simple way, pointing out the advantages of
a TB model with respect to first-principles calculations.
The explicit analysis of this issue is discussed in
Section \ref{s:param}.
We anticipate here that the effects of the off-diagonal spin-flip
terms result to be negligible for all the cases of interest here.
{\em This is essentially due to the fact that such processes
involve virtual transitions towards high-order energy states.}\cite{OR13}
At a very high degree of accuracy, we  are thus justified
in neglecting the spin-flip terms and retaining
in (\ref{Eq:HSO})
only the
spin-conserving terms $\propto \lambda_a\hat{L}_a^z \hat{S}_a^z$.
An immediate consequence of that is that
the {\it even} and {\it odd} sectors of the Hamiltonian remain
uncoupled,
allowing us to restrict our analysis, for the low-energy states
of the valence and conduction bands, only to the E sector.

\subsection{Bulk case}

Once introduced the TB model for a single-layer
in the presence of SOC, it is quite straightforward
to construct a corresponding theory for the bulk and bilayer
systems by including the relevant
inter-layer hopping terms in the Hamiltonian. Considering that the unit cell is now doubled,
we can thus write the Hamiltonian for bulk {\em MX}$_2$
in the presence of
SOC in the matrix form:
\begin{eqnarray}
\hat{H}_{\rm Bulk}({\bf k})
&=&
\hat{H}_{\rm Bulk}^{\rm sl}({\bf k})
\otimes {\mathbb 1}_2
+
\hat{H}_{\rm Bulk}^{\rm SO},
\label{eqB}
\end{eqnarray}
which is a $44\times44$ matrix due to the doubling of the unit cell with respect to the single-layer case discussed in Sec. \ref{Sec:SL}.

Here $\hat{H}_{\rm Bulk}^{\rm sl}({\bf k})$
represents the spinless Hamiltonian for the bulk system,
\begin{eqnarray}
\hat{H}_{\rm Bulk}^{\rm sl}({\bf k})
&=&
\left(
\begin{array}{cc}
\hat{H}_1^{\rm sl} &  \hat{H}_{\perp,\rm Bulk} \\
\hat{H}_{\perp,\rm Bulk}^\dagger &
\hat{H}_2^{\rm sl} 
\end{array}
\right),
\label{Eq:HBulk}
\end{eqnarray}
where
$\hat{H}_i^{\rm sl}$ describes the
spinless Hamiltonian (i.e. in the absence of SOC) for the layer $i=1,2$,
while
$\hat{H}_{\perp,\rm Bulk}$ accounts for the $11 \times 11$
Hamiltonian describing interlayer hopping between $X$ atoms beloging
to different layers.
We remind that $\hat{H}_2^{\rm sl}$ is related to $\hat{H}_1^{\rm sl}$
through the following relation dictated by 
the lattice structure:\cite{CG13}
 \be
H_{2,\alpha,\beta}^{\rm sl}(k_x,k_y)
=
P_{\alpha}P_{\beta}H_{1,\alpha,\beta}^{\rm sl}(k_x,-k_y),
\ee
where $P_{\alpha}=+(-)1$ if the orbital $\alpha$ has even (odd)
symmetry with respect to $y\rightarrow -y$.
Furthermore, the (spin-diagonal) interlayer term
$\hat{H}_{\perp,\rm Bulk}$ can be written as:
\begin{eqnarray}
\hat{H}_{\perp,\rm Bulk} ({\bf k})
&=&
\left(
\begin{array}{cc}
\hat{I}_{\rm E}\cos \zeta &  \hat{I}_{\rm EO}\sin \zeta \\
-\hat{I}^{\rm T}_{\rm EO}\sin \zeta & \hat{I}_{\rm O}\cos \zeta 
\end{array}
\right),
\label{Eq:Hperp}
\end{eqnarray}
where $\zeta=k_zc/2$
($c$ being the vertical size of the unit cell in the bulk system),
and where the matrices $\hat{I}_{\rm E}$, $\hat{I}_{\rm O}$ and
$\hat{I}_{\rm  EO}$ describe
the inter-layer hopping between the $p$ orbitals of
the adjacent chalcogen atoms.
One can notice that interlayer hopping leads, for an arbitrary wave-vector $\bk$, to a mixture of the E and O sectors of the Hamiltonian, which is accounted for by the term $\hat{I}_{\rm EO}$ in (\ref{Eq:Hperp}).\cite{CG13}
The analysis is however simplified at specific high-symmetry points
of the BZ, as we discuss below.\footnote{See Ref. \onlinecite{CG13} for an explicit expression of
all the matrix elements of the Hamiltonian
(\ref{Eq:HBulk}).}

Finally $\hat{H}_{\rm Bulk}^{\rm SO}$ in Eq. (\ref{eqB}) accounts
for the spin-orbit coupling in the bulk system,
and it can be written as:
\begin{eqnarray}
\hat{H}_{\rm Bulk}^{\rm SO}
&=&
\left(
\begin{array}{cccc}
\hat{M}^{\uparrow\uparrow}
&  0
& \hat{M}^{\uparrow\downarrow} & 0\\
0 &
\hat{M}^{\uparrow\uparrow}
& 0 & \hat{M}^{\uparrow\downarrow}\\
\hat{M}^{\downarrow\uparrow} & 0 &
 \hat{M}^{\downarrow\downarrow}&
0
\\
0 & \hat{M}^{\downarrow\uparrow} & 
0 &
\hat{M}^{\downarrow\downarrow}
\end{array}
\right),
\label{Eq:SOBulk}
\end{eqnarray}
where both the spin-diagonal ($\hat{M}^{\sigma\sigma}$)
and spin-flip ($\hat{M}^{\sigma\bar{\sigma}}$)
processes induced by the atomic spin-orbit interaction are present.

Eqs. (\ref{eqB})-(\ref{Eq:SOBulk}) provide the general basic framework for a
deeper analysis in more specific cases.
In particular, as already mentioned above, the spin-flip terms
triggered by SOC can be substantially neglected for
all the cases of interest without loosing accuracy.
The total Hamiltonian (\ref{eqB}) can thus be divided in two $22 \times 22$
blocks $\hat{H}_{\rm Bulk}^{\sigma\sigma}({\bf k})$ related by the
symmetry $\hat{H}_{\rm Bulk}^{\uparrow\uparrow}({\bf k})=
\hat{H}_{\rm Bulk}^{\downarrow\downarrow}(-{\bf k})$.
Further simplifications are available at specific symmetry points
of the BZ.
More specifically,
we can notice that for $k_z=0$
the E and O sectors remain uncoupled.
Focusing, at low-energies for the conduction and valence bands,
only on the E sector, we can write
\begin{eqnarray}
\hat{H}_{\rm Bulk,E}({\bf k},k_z=0)
&=&
\hat{H}_{\rm Bulk,E}^{\rm sl}({\bf k})
+
\hat{H}_{\rm Bulk,E}^{\rm SO},
\label{eqBE}
\end{eqnarray}
where
\begin{eqnarray}
\hat{H}_{\rm Bulk,E}^{\rm sl}({\bf k})
&=&
\left(
\begin{array}{cccc}
\hat{H}_{\rm E,1}
&  \hat{I}_{\rm E}
& 0 & 0\\
\hat{I}_{\rm E}^\dagger &
\hat{H}_{\rm E,2}
& 0 & 0\\
0 & 0 &
\hat{H}_{\rm E,1}&
\hat{I}_{\rm E}
\\
0 & 0& 
\hat{I}_{\rm E}^\dagger &
\hat{H}_{\rm E,2}
\end{array}
\right),
\end{eqnarray}
and
\begin{eqnarray}
\hat{H}_{\rm Bulk,E}^{\rm SO}
&=&
\left(
\begin{array}{cccc}
\hat{M}_{\rm EE}^{\uparrow\uparrow}
&  0
&0 & 0\\
0 &
\hat{M}_{\rm EE}^{\uparrow\uparrow}
& 0 & 0\\
0 & 0 &
\hat{M}_{\rm EE}^{\downarrow\downarrow}&
0
\\
0 & 0 & 
0 &
\hat{M}_{\rm EE}^{\downarrow\downarrow}
\end{array}
\right),
\label{sobulkkk}
\end{eqnarray}
where
the explicit expression of each block Hamiltonian
is also reported in Appendix \ref{App:SOC}.

\subsection{Bilayer}

The Hamiltonian for the bilayer can also be derived in a very similar
form as  in the bulk case.
In particular, we can write:
\begin{eqnarray}
\hat{H}_{\rm 2L}({\bf k})
&=&
\hat{H}_{\rm 2L}^{\rm sl}({\bf k})
+
\hat{H}_{\rm 2L}^{\rm SO}.
\label{Ep:BL}
\end{eqnarray}
Since we are considering intrinsic SOC, 
thus it is not affected by the interlayer coupling.
Therefore we have $\hat{H}_{\rm 2L}^{\rm SO}
=\hat{H}_{\rm Bulk}^{\rm SO}$,
where $\hat{H}_{\rm Bulk}^{\rm SO}$ is defined in
Eq. (\ref{sobulkkk}).

On the other hand,
similar to the bulk case in Eq. (\ref{Eq:HBulk}),
the spinless tight-binding
term $\hat{H}_{\rm 2L}^{\rm sl}({\bf k})$
for the bilayer case can be written as:
\begin{eqnarray}
\hat{H}_{\rm 2L}^{\rm sl}({\bf k})
&=&
\left(
\begin{array}{cc}
\hat{H}_1^{\rm sl} &  \hat{H}_{\perp,\rm 2L} \\
\hat{H}_{\perp,\rm 2L}^\dagger &
\hat{H}_2^{\rm sl} 
\end{array}
\right),
\label{Eq:H2L}
\end{eqnarray}
where now
\begin{eqnarray}
\hat{H}_{\perp,\rm 2L} ({\bf k})
&=&
\frac{1}{2}\left(
\begin{array}{cc}
\hat{I}_{\rm E} &  \hat{I}_{\rm EO} \\
-\hat{I}^{\rm T}_{\rm EO} & \hat{I}_{\rm O}
\end{array}
\right).
\label{Eq:HperpBL}
\end{eqnarray}
Note that Eq. (\ref{Eq:HperpBL}) can be obtained as
limiting case of Eq. (\ref{Eq:Hperp}) by setting $\zeta=\pi/4$,
corresponding to the effective uncoupling of bilayer blocks.

\section{Tight-binding parameters and comparison with DFT
  calculations}
\label{s:param}

After having developed a suitable tight-binding model for single and
multi-layer {\em MX}$_2$ compounds, we compare in this section the band structure obtained by the TB model to the corresponding band structure obtained from DFT methods. 
An appropriate set of tight-binding parameters can be derived by fitting the low-energy dispersion of
the conduction and valence bands of these compounds in the whole
BZ, including the secondary minimum of the conduction band
at the Q point, along the $\Gamma$-K line. The crystal field
$\Delta_1$ is obtained by fixing the minimum at K
of the electronic bands belonging to the {\it odd} block to the same
energy of the DFT calculations.
The only left unknown parameters are thus the
atomic spin-orbit constants $\lambda_{M}$ and $\lambda_X$ for
the transition metal and for the chalcogen atom, respectively.
We take the corresponding values from Ref. \onlinecite{LX13} and \onlinecite{KGF13},
and we list the full set of TB parameters for MoS$_2$ and WS$_2$
in Table \ref{Tab:TB}.
%%%%%
\begin{table}[t]
\begin{tabular}{lclcrcr}
\hline
\hline
                          &                                &                   &                                 &    MoS$_2$     & \hspace{0.5truecm} & WS$_2$ \\
\hline
\\
  SOC   & \hspace{0.5truecm} &$\lambda_{\rm Mo}$ & \hspace{0.5truecm} &  0.075 & \hspace{0.5truecm} & 0.215 \\
                        & \hspace{0.5truecm} &$\lambda_{\rm S}$ & \hspace{0.5truecm} &  0.052 & \hspace{0.5truecm} & 0.057 \\
\\
   Crystal Fields & \hspace{0.5truecm} &$\Delta_0$ & \hspace{0.5truecm} &  -1.512 & \hspace{0.5truecm} & -1.550 \\
                          & \hspace{0.5truecm} &$\Delta_1$ & \hspace{0.5truecm} &  0.419         & \hspace{0.5truecm} & 0.851 \\
                          & \hspace{0.5truecm} &$\Delta_2$ & \hspace{0.5truecm} & -3.025  & \hspace{0.5truecm} & -3.090  \\
                          & \hspace{0.5truecm} &$\Delta_p$ & \hspace{0.5truecm} &  -1.276 & \hspace{0.5truecm} & -1.176 \\
                          & \hspace{0.5truecm} &$\Delta_z$ & \hspace{0.5truecm} &  -8.236 & \hspace{0.5truecm} & -7.836  \\
\\                
Intralayer Mo-S & \hspace{0.5truecm} &$V_{pd\sigma}$ & \hspace{0.5truecm} &  -2.619 & \hspace{0.5truecm} & -2.619 \\
                          & \hspace{0.5truecm} &$V_{pd\pi}$ & \hspace{0.5truecm} &  -1.396  & \hspace{0.5truecm} & -1.396 \\
\\                      
Intralayer Mo-Mo& \hspace{0.5truecm} &$V_{dd\sigma}$ & \hspace{0.5truecm} &  -0.933 & \hspace{0.5truecm} &  -0.983\\
                          & \hspace{0.5truecm} &$V_{dd\pi}$ & \hspace{0.5truecm} &  -0.478 & \hspace{0.5truecm} &  -0.478\\
                          & \hspace{0.5truecm} &$V_{dd\delta}$ & \hspace{0.5truecm} &  -0.442 & \hspace{0.5truecm} &  -0.442\\
\\                        
Intralayer S-S & \hspace{0.5truecm} &$V_{pp\sigma}$ & \hspace{0.5truecm} &  0.696 & \hspace{0.5truecm} &  0.696\\
                          & \hspace{0.5truecm} &$V_{pp\pi}$ & \hspace{0.5truecm} &  0.278 & \hspace{0.5truecm} &  0.278 \\
\\                         
Interlayer S-S & \hspace{0.5truecm} &$U_{pp\sigma}$ & \hspace{0.5truecm} &  -0.774 & \hspace{0.5truecm} &  -0.774\\
                          & \hspace{0.5truecm} &$U_{pp\pi}$ & \hspace{0.5truecm} &  0.123 & \hspace{0.5truecm} &  0.123\\
\hline
\hline
\end{tabular}
\caption{Spin-orbit coupling $\lambda_{\alpha}$ and tight-binding parameters for single-layer MoS$_2$ and WS$_2$
($\Delta_\alpha$, $V_\alpha$)
as obtained by fitting the low energy conduction and valence bands.
Also shown are the inter-layer hopping parameters $U_\alpha$
relevant for bulk compounds. The Slater-Koster parameters for MoS$_2$ are taken from Ref. \onlinecite{CG13}, and the SOC terms from Ref. \onlinecite{LX13} and \onlinecite{KGF13}. All hopping terms $V_\alpha$, $U_\alpha$,
crystal fields $\Delta_\alpha$, and spin-orbit coupling $\lambda_a$
are in units of eV.}
\label{Tab:TB}
\end{table}
%%%%%%
Therefore, we can compare the resulting band structure for the full tight-binding
model in the presence of SOC, with corresponding first-principles results
including also spin-orbit interaction.

DFT calculations were performed
using the \textsc{Siesta} code.\cite{soler,artacho}
The spin-orbit interaction is treated as in Ref.  \onlinecite{seivane}.
We use the exchange-correlation potential of
Ceperley-Alder\cite{ceperly} as parametrized by Perdew and Zunger.\cite{perdew}
We use also a split-valence double-$\zeta$ basis set including
polarization functions.\cite{artacho2} The energy cutoff and the Brillouin
zone sampling were chosen to converge the total energy.
Lattice parameters for MoS$_2$ and WS$_2$
were chosen according to their experimental values,
as reported in Refs. \onlinecite{BMY72} and \onlinecite{schutte},
and they are listed in Table \ref{t:DFTvalues}. 
\begin{table}[b]
\begin{tabular}{|l|c|c|c|}
\hline
\hline
 & $a$ & $u$ & $c'$ \\
\hline
MoS$_2$ 1L & $3.16$ & $1.586$ & $-$ \\
\hline
MoS$_2$ 2L & $3.16$ & $1.586$ & $6.14$  \\
\hline
MoS$_2$ Bulk & $3.16$ & $1.586$ & $6.14$  \\
\hline
WS$_2$ 1L & $3.153$ & $1.571$ & $-$ \\
\hline
WS$_2$ 2L & $3.153$ & $1.571$ & $6.1615$  \\
\hline
WS$_2$ Bulk & $3.153$ & $1.571$ & $6.1615$  \\
\hline
\hline
\end{tabular}
\caption{
Lattice parameters used for DFT calculation for single-layer,
bilayer and bulk MoS$_2$ and WS$_2$ systems, as taken from
Refs. \onlinecite{BMY72} and \onlinecite{schutte}, respectively. $a$ represents the $M$-$M$ atomic distance,
$u$ the internal vertical distance between the $M$ plane
and the $X$ plane, and $c'$ the distance between the $M$ layers.
In bulk systems the $z$-axis lattice parameter is given by $c=2c'$.
All values are in \AA\, units.
}
\label{t:DFTvalues}
\end{table}

\begin{figure}[t]
\begin{center}
\mbox{
\includegraphics[width=0.5\columnwidth]{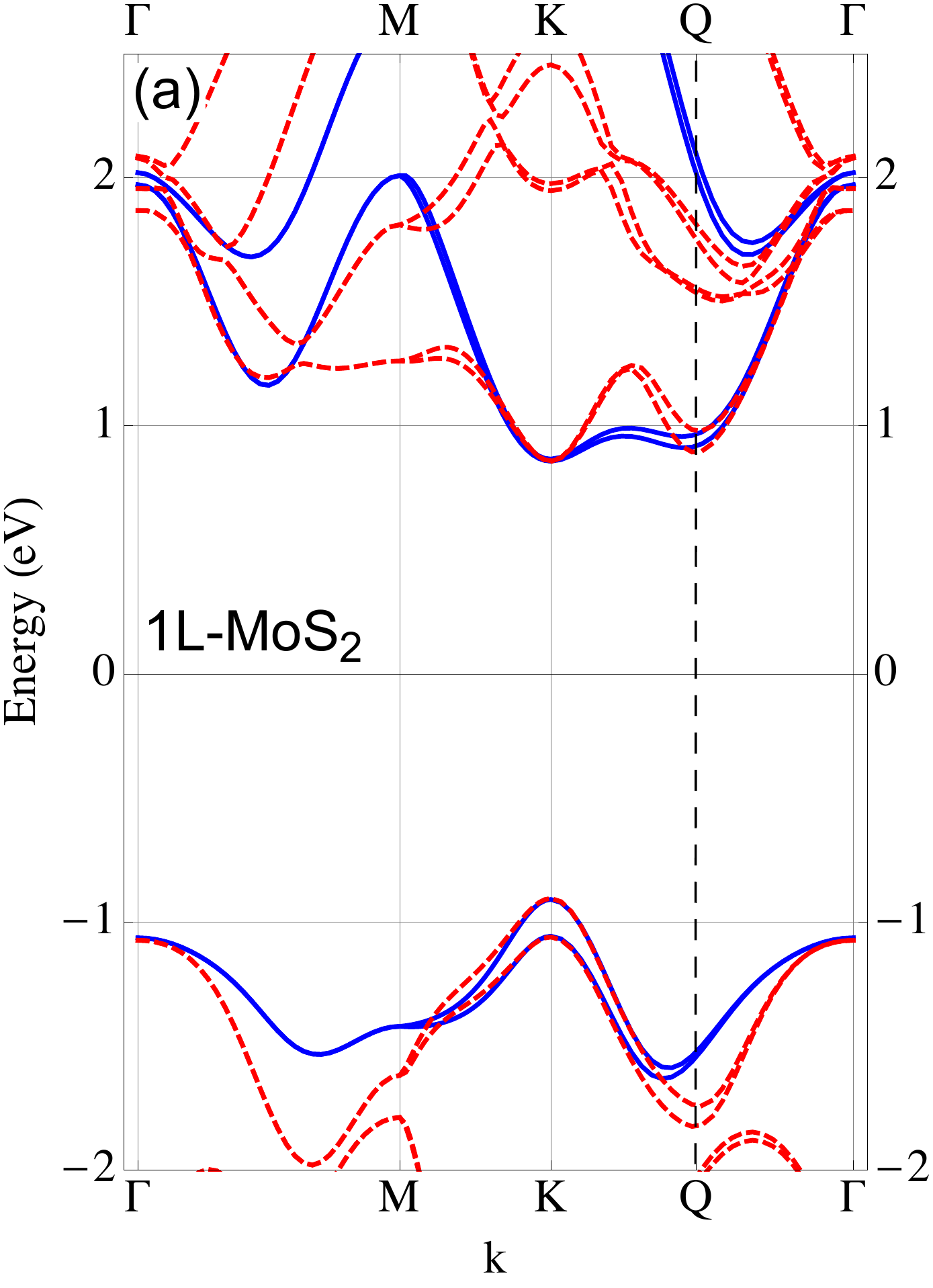}
\includegraphics[width=0.5\columnwidth]{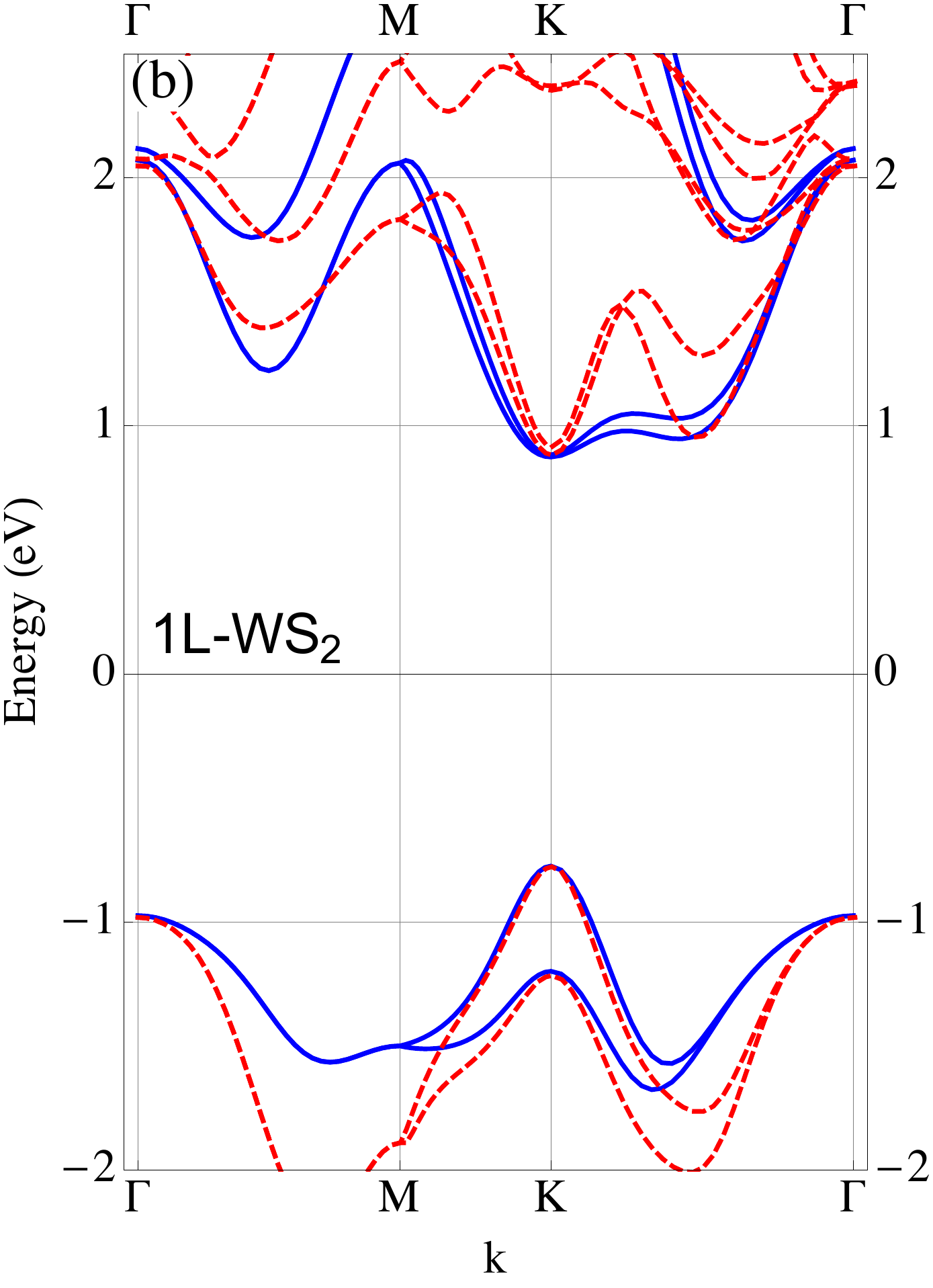}
} 
\mbox{
\includegraphics[width=0.5\columnwidth]{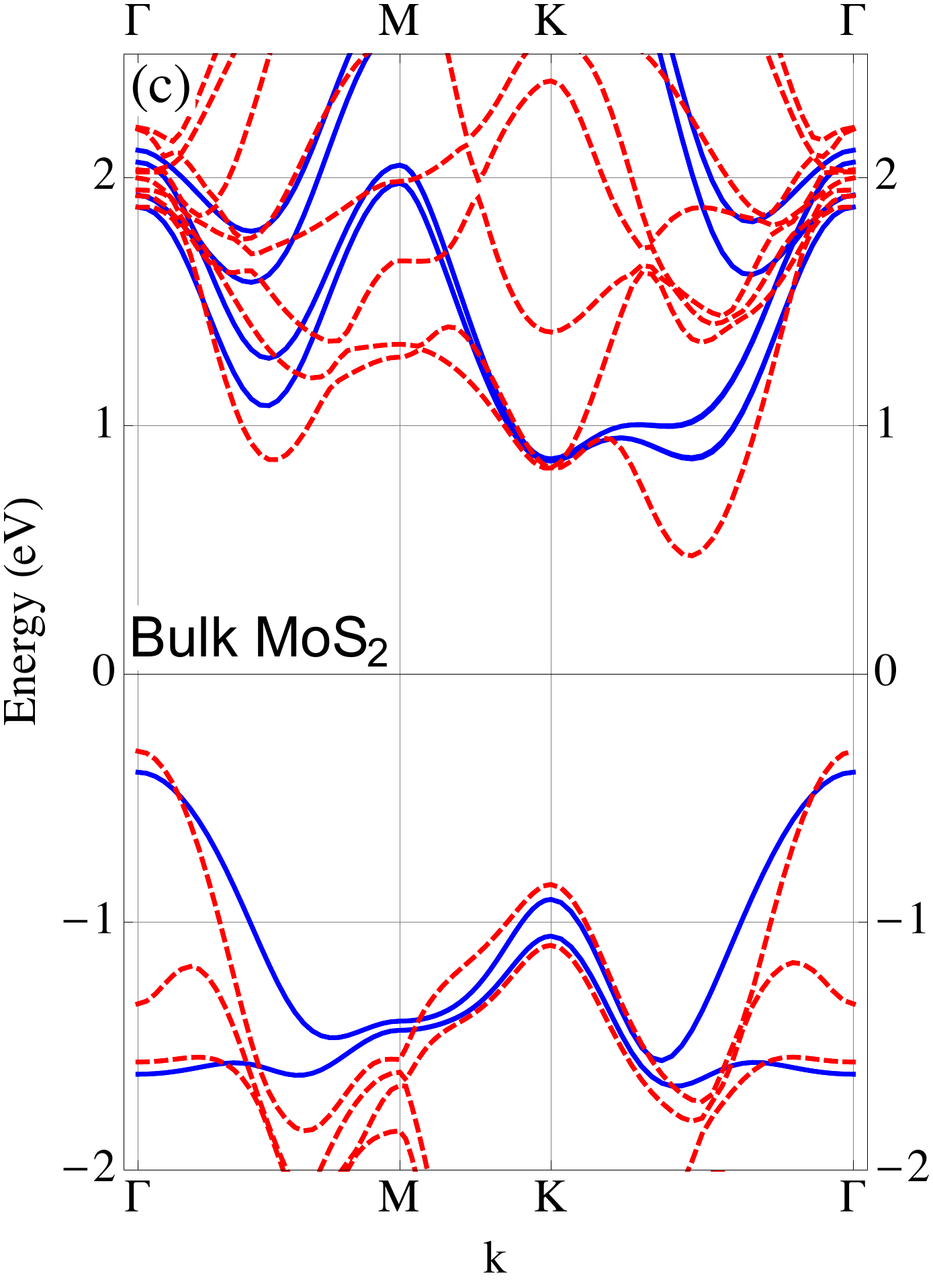}
\includegraphics[width=0.5\columnwidth]{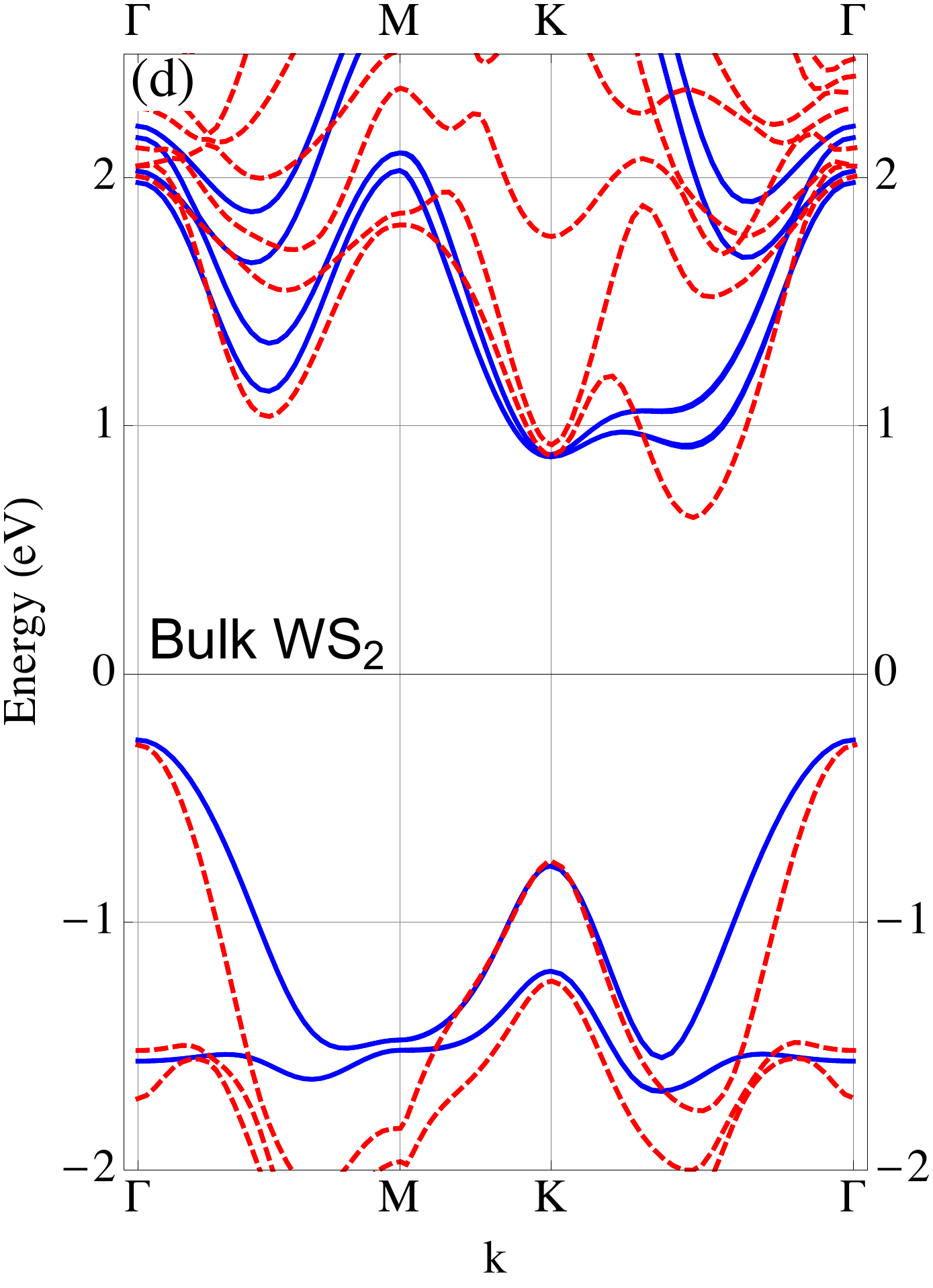}
}
\end{center}
\caption{Band structure of single-layer and bulk MoS$_2$ and WS$_2$
including SO interaction.
Red dashed lines correspond to DFT calculations
and solid blue lines to tight-binding calculations using the sets of parameters
given in Table \ref{Tab:TB}. The vertical dashed line in panel (a) indicates the position of the minimum of the conduction band, referred in the text as Q.}
\label{Fig:BandsSO11x11}
\end{figure}

The representative band structure for monolayer MoS$_2$ and WS$_2$,
as well as for the bulk counterpart, are shown in
Fig. \ref{Fig:BandsSO11x11},
for both DFT (dashed red lines) and TB calculations (solid blue lines). We observe that the TB model with the set of Slater-Koster parameters provided in Table \ref{Tab:TB} leads to a reasonable fitting of the DFT band structure. In particular we see that, for single layer samples [panels (a) and (b)] the edges of the valence band at K and $\Gamma$, as well as the edges of the conduction band at K and Q [which position is marked by a vertical dash in Fig. \ref{Fig:BandsSO11x11}(a)] are properly captured by the TB model.  The TB valence bands are less dispersive than the DFT bands in the intermediate regions between high symmetry points.  The experimental bands measured by ARPES also seem to be flatter than the DFT bands, as it has been recently shown in Ref. \onlinecite{JO13}.  However it is important to notice that those experimental results for the band dispersion can be affected by the interaction between the $MX_2$ crystals and the substrate, which is not consider neither in DFT calculations nor in the TB model. The TB band structure for bulk samples, shown in Fig. \ref{Fig:BandsSO11x11}(c) and (d), have been obtained by adding {\it only} two extra Slater-Koster parameters, $U_{pp\sigma}$ and $U_{pp\pi}$, which account for inter-layer hopping between $p$ orbitals of the adjacent chalcogen atoms of different layers. The obtained band structure for the valence band reproduce reasonably well the DFT band structure, as well as the experimental band structure measured by ARPES,\cite{JO13} and accounts for the direct- to indirect-gap transition when going from 1L to bulk materials.\cite{CG13} As for the conduction band, the minimum at K is also captured by the TB model, but the energy of the minimum at Q does not agree with DFT results. The inclusion of hopping terms between $M$ orbitals of different layers, as well as next nearest neighbor hopping terms, could improve such fitting.\cite{footnote} 
 However, we notice that no experimental measurements of the conduction band dispersion are available so far in the literature that could serve to validate the DFT and the TB results presented here.

\begin{figure}[b]
\begin{center}
\mbox{
\includegraphics[width=0.5\columnwidth]{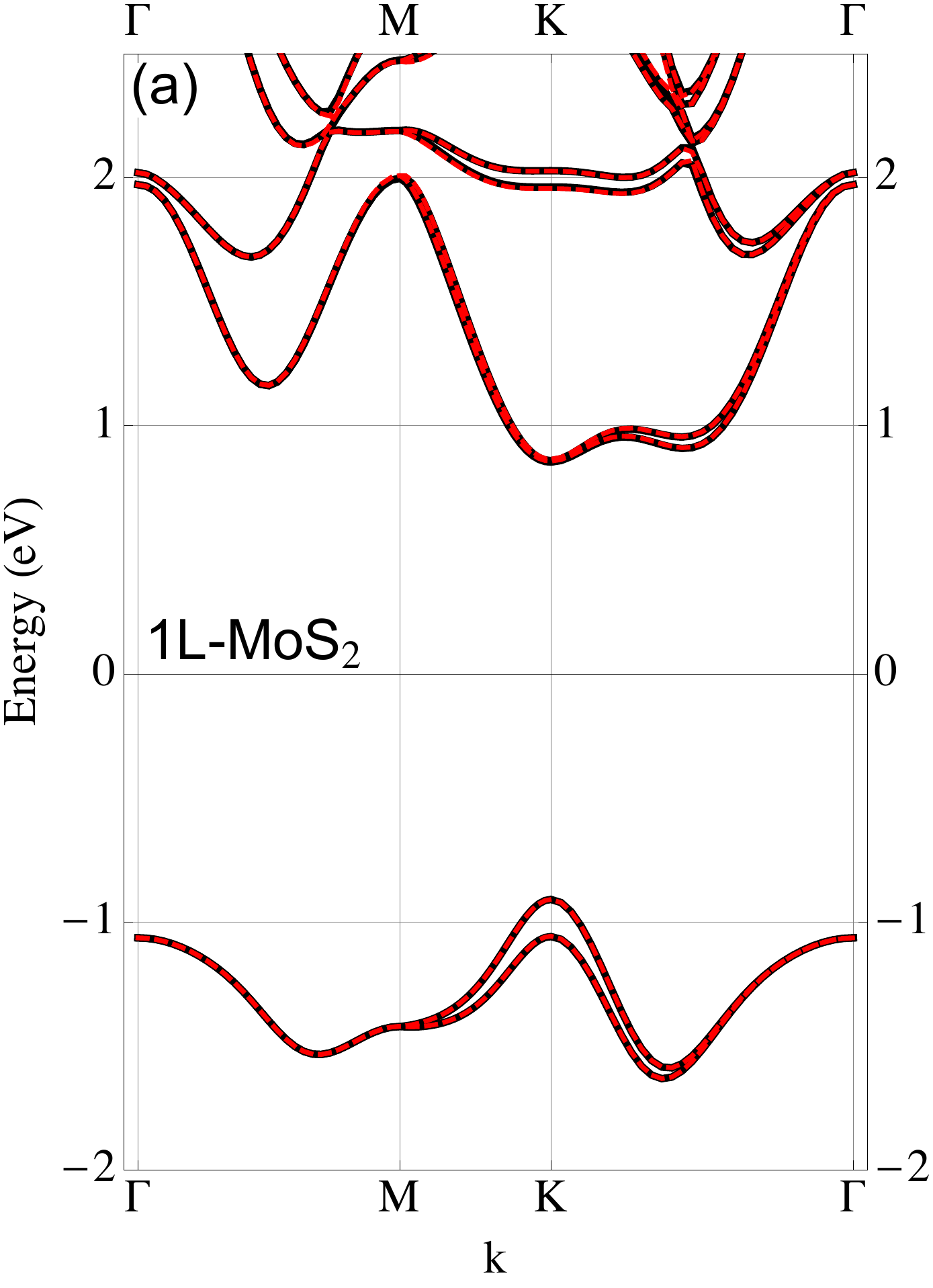}
\includegraphics[width=0.5\columnwidth]{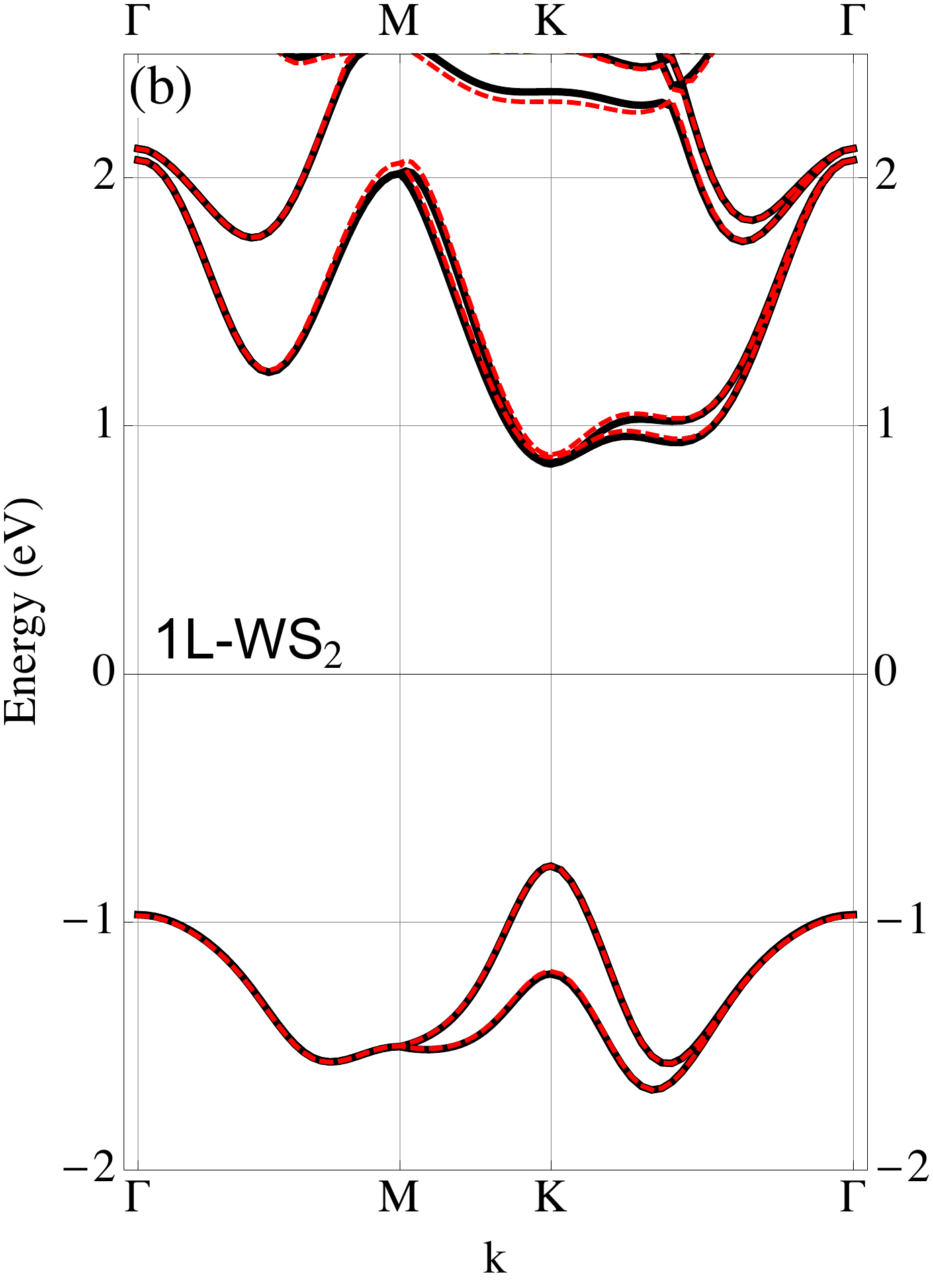}
} 
\end{center}
\caption{Tight-binding band structure of single-layer MoS$_2$ (a) and WS$_2$ (b) including SO
  interaction. Red dashed lines corresponds to the TB bands including
  the whole SO coupling terms. Black solid lines correspond to the TB
  band structure including only the spin-conserving terms of the SO
  coupling.  
  }
\label{Fig:BandsSO-NoSF}
\end{figure}

In addition to the above remarks,
a fundamental advantage of the TB model with respect to
first-principles calculations is that it permits to investigate,
in an analytical way, the relevance of the microscopic underlying
processes.
We have already mentioned above how transverse spin-flip fluctuations
play here a marginal role and they can be disregarded, making
the overall modeling of the spin-orbit interaction extremely
direct and simple.
We can now explicitly address and quantify this issue by
comparing in the TB model the band structures
obtained by using the full SOC as described by Eq. (\ref{Eq:HSO})
and the one obtained considering only the last spin-diagonal terms 
$\hat{L}_a^z\hat{S}_a^z$. The results are shown in Fig. \ref{Fig:BandsSO-NoSF}
where we compare, for single-layer MoS$_2$ and WS$_2$, the total band structure (red dashed lines)
obtained by considering the full spin-orbit interaction (\ref{Eq:HSO})
with the one obtained using the spin-conserving part
[third term in Eq. (\ref{Eq:HSO})].
As we can see in Fig. \ref{Fig:BandsSO-NoSF}(a) there is an almost perfect overlapping of the
band structures for MoS$_2$ obtained including and neglecting the spin-flip terms,
demonstrating  the negligible role of these processes and the validity of the approximation. The effect is still weak but more noticeable for the case of WS$_2$ [Fig. \ref{Fig:BandsSO-NoSF}(b)], due to the larger atomic SOC associated to the heavier W atoms, as compared to Mo. 

Of special interest is the minimum of the conduction band at the K point of the BZ: here, as discussed in Refs. \onlinecite{LX13,KB13,KGF13}, the competition between second order spin-flip processes associated with the transition metal atom $M$ and first order (spin-conserving) processes of the chalcogen atom $X$, are responsible for the crossing/non crossing of the conduction bands in a very narrow region close to this K point.

\section{Discussion}
\label{Sec:Discussion}

The TB model introduced in Sec. \ref{s:model},
for single-layer and multi-layer compounds,
and the specific Slater-Koster parameter set discussed in Sec. \ref{s:param}
provide a comprehensive tool for the study of the electronic properties
and the entanglement between different degrees of freedom
(spin, orbital, valley, layer, lattice) in these compounds
in the presence of a relevant SOC acting
both on the chalcogen $X$ and on the transition metal atoms $M$.
As we summarize in the present Section, such physics results
to be relevant not only for the valence bands,
whose band edge in the single layer materials is mainly built
by the $M$ orbitals $d_{xy}$, $d_{x^2-y^2}$,
but also for the conduction band and for the secondary extrema
of both conduction and valence bands, whose energy can be effectively
tuned by the interlayer coupling and by the spin-orbit interaction itself.

\subsection{Spin-polarized pockets in the Fermi surfaces}
\label{Sec:FS}

The role of the SOC on the spin-orbital-valley
entanglement at the band edge at K 
of the single-layer and bilayer compounds has been
previously discussed in the literature, using mainly low-energy
effective Hamiltonians focused on the role of the transition metal
$M$ $d$-orbitals and of their corresponding spin-orbit coupling.\cite{XY12,RMA13,LZN13,KL13,WW14,COG13,GMP14}
Such scenario can be now well reproduced by the present TB model
and generalized to the whole BZ.

\begin{figure*}[t]
\begin{center}
\mbox{
\includegraphics[width=0.75\columnwidth]{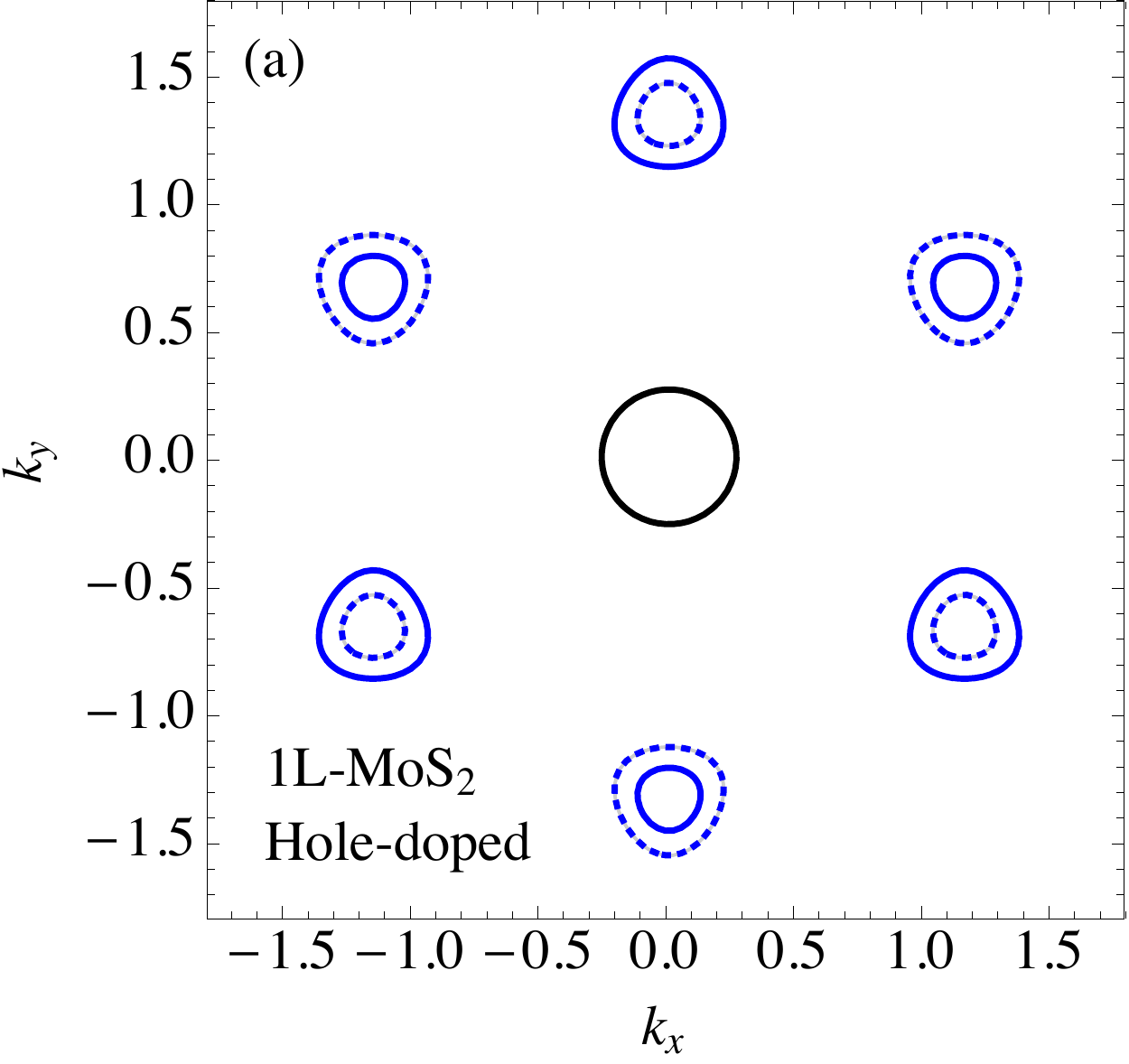}
\includegraphics[width=0.75\columnwidth]{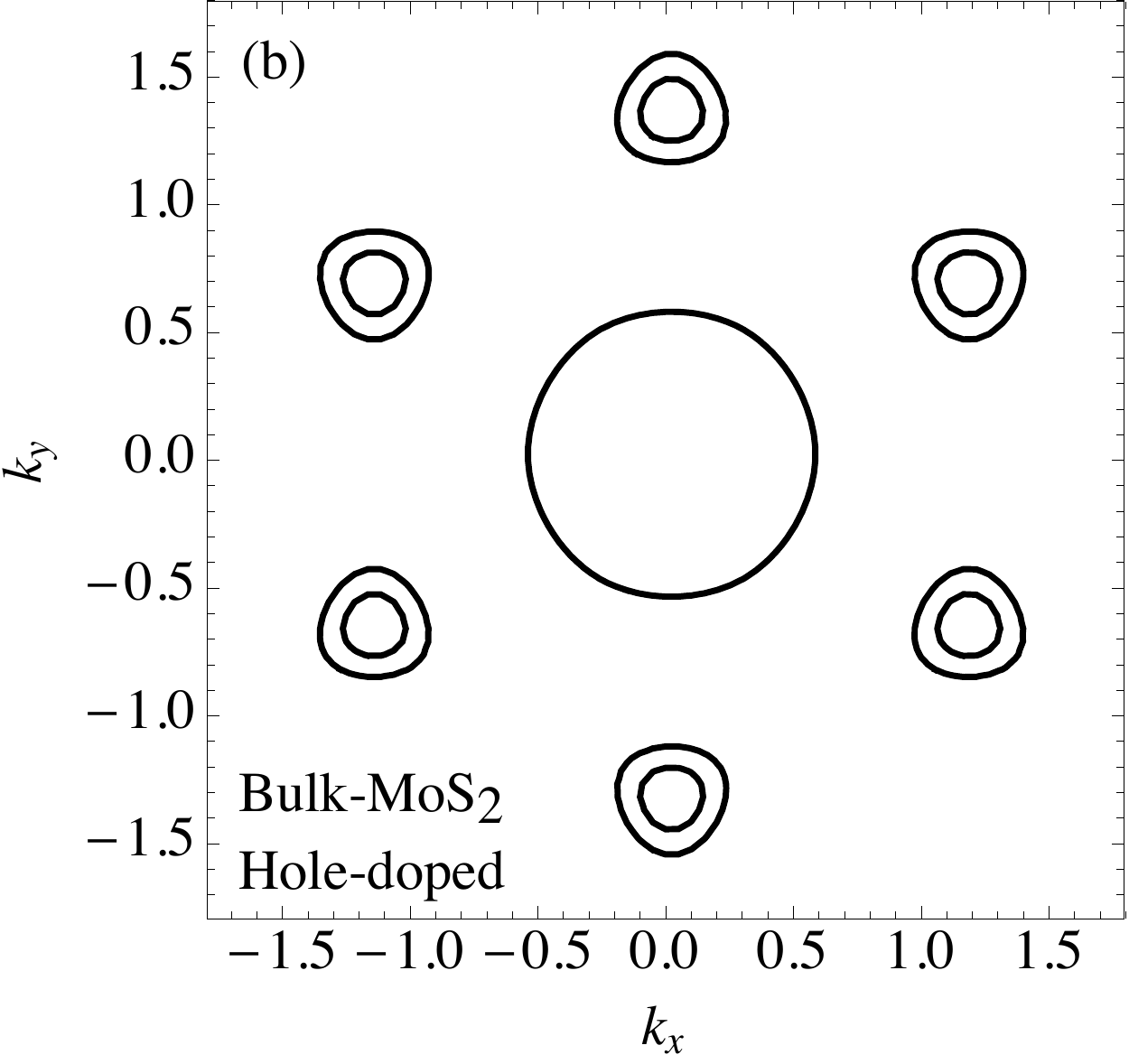}
} 
\mbox{
\includegraphics[width=0.75\columnwidth]{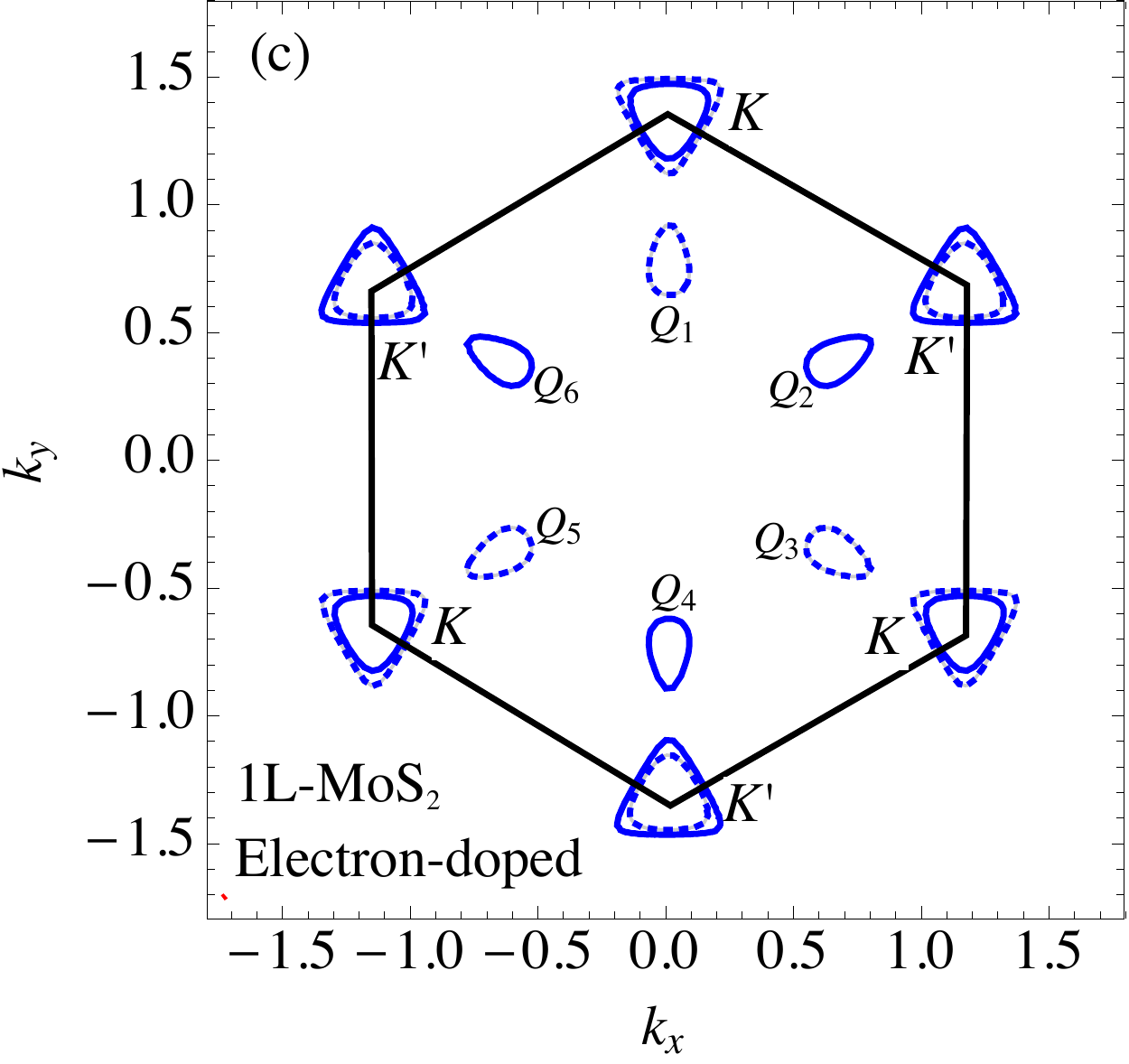}
\includegraphics[width=0.75\columnwidth]{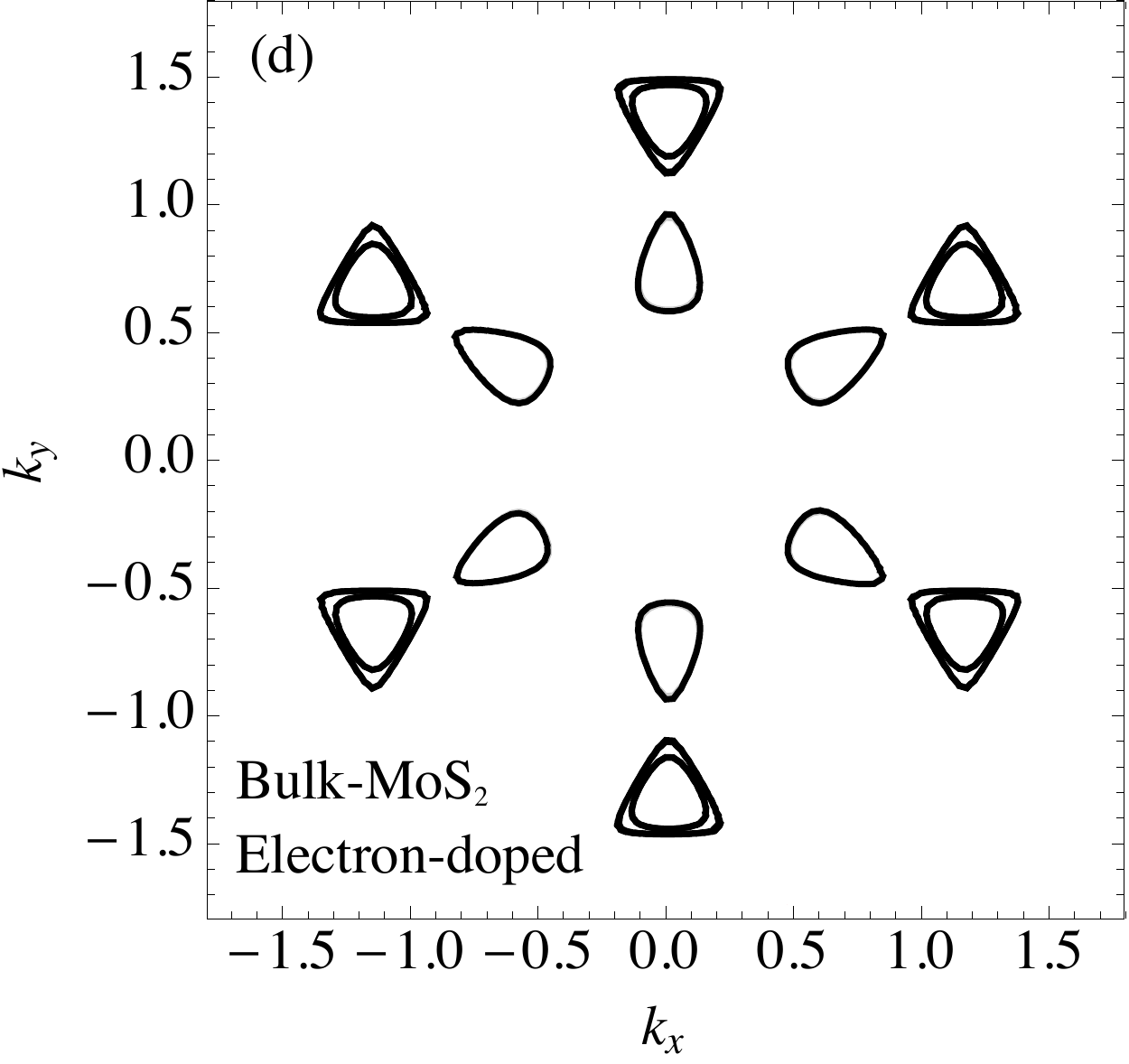}
} 
\end{center}
\caption{Fermi surfaces of MoS$_2$, obtained from the TB band structure. 
Panels (a) and (c) correspond to single-layer and panels (b) and (d) to the bulk.
Top panels represent hole-doped systems,
with the Fermi energy  in the valence band (at $E_F=-1.134$~eV),
whereas bottom panels represent electron-doped systems,
with the Fermi energy in the conduction band ($E_F=0.95$~eV).
Energies are measured with respect to the zero of the TB Hamiltonian.
The hexagonal 2D BZ is shown in (c) by the black solid lines.
In panels (a) and (c), solid blue and dashed blue lines correspond to Fermi surfaces with main
$S_z=\uparrow$  and $S_z=\downarrow$ polarization, respectively. Solid black lines indicate pockets which are degenerate in spin,
like the central pocket in (a) (around the $\Gamma$ point) and all the pockets in the Fermi surfaces of the bulk compound [panels (b) and (d)].
}
\label{Fig:FS}
\end{figure*}

The spin-orbit coupling, in particular, is expected to be most
relevant for the band edges of the valence band at the K point,
whose orbital content is mainly associated with the
$d_{xy}$ and $d_{x^2-y^2}$ orbitals of the transition metal.
A large band splitting induced by the SOC is thus predicted in this
case.
Such feature is indeed well captured by the TB model.
In Fig. \ref{Fig:FS}(a) and (b) we show
the Fermi surfaces obtained with the present TB model,
including atomic SOC, for a finite hole-doping probing
the valence band of both single-layer and bulk compounds.
In order to point out the different physics occurring close to the different
band edges at K and $\Gamma$ points, we show here
Fermi surfaces corresponding to a sizable negative Fermi energy
cutting both edges at K and $\Gamma$.
In particular, the central Fermi pocket located
around $\Gamma$ appears to be spin degenerate,
for both single-layer and bulk systems
since its orbital character is mainly due to the
$d_{3z^2-r^2}$ orbitals of $M$ and to the $p_z$ orbitals of $X$,\cite{CG13}
both of them with $L_z=0$.
On the other hand, the pockets around K and K' are mainly due to
the $d_{x^2-y^2}$ and $d_{xy}$ orbitals of the metal $M$ (with
$|m|=2$), plus a minor component of $p_x$ and $p_y$ orbitals
of the chalcogen $X$ (with $|m|=1$).
This results in a finite SOC splitting of the valence band at the K
and K' points,
due mainly to first order spin-orbit coupling on
the $d$ orbitals of $M$.
Furthermore, because of the lack of inversion symmetry in single layer
samples (or in multi-layer samples with an odd number of layers),
the spin degeneracy is lifted, presenting an opposite spin polarization
on different valleys.\cite{XY12}
This feature is well reproduced by our model and shown
in Fig. \ref{Fig:FS}(a), where Fermi surfaces with main $S_z=\uparrow$
character are denoted by solid blue lines,
while Fermi surfaces with main $S_z=\downarrow$
character are denoted by dashed blue lines.
On the other hand,
the Fermi surfaces of hole-doped bulk MoS$_2$, for the same $E_F$,
are shown in Fig. \ref{Fig:FS}(b).
Since the maximum of the valence band for the bulk compound,
because of the interlayer coupling,
is located at the $\Gamma$ point
 [see the band structure of Fig. \ref{Fig:BandsSO11x11}(c)],
the central pocket in Fig. \ref{Fig:FS}(b)
is considerably larger than in Fig. \ref{Fig:FS}(a) for single layer
samples.
The the double Fermi surfaces around the K and K' points in
\ref{Fig:FS}(b) are spin degenerate, as impose by inversion symmetry. A recent set of ARPES measurements for MoS$_2$ and MoSe$_2$\cite{AH13} have shown the importance of the SOC in the band structure, obtaining experimental constant energy contours in very much agreement with those presented in Fig. \ref{Fig:FS}(a) and (b).

Although smaller and less noticed,\cite{OR13,CL12,Kosmider_etal_2013,KF13,KB13,KGF13} 
a spin-valley coupling is present also for the conduction band edge
of the single-layer systems at the
K and K' points.
It is important to remind here that the orbital character in
these points of the BZ is mainly associated
with the $d_{3z^2-r^2}$ orbital (with $m=0$) of the transition metal $M$,
but with a finite contribution from the
$p_x$ and $p_y$ orbitals of the chalcogen, with $m=\pm 1$).\cite{CG13}
The spin-orbit coupling of the chalcogen atom $X$, mainly through the
diagonal term $L^z_X S^z_X$, results thus in a smaller but finite
splitting of the conduction band edge, as it can be also inferred
by the Fermi surfaces for electron-doped single-layer compounds,
as shown in Fig. \ref{Fig:FS}(c).
It is worth to stress that, although the resulting spin-induced
splitting can be quite small, the entanglement between band splitting,
spin and valley degrees appears to be quite strong, so that the lower
band is $\uparrow$ polarized and the upper band $\downarrow$ polarized (or viceversa,
depending on the valley).
Note also that, 
although the atomic spin-orbit coupling due to the sulfur
in MoS$_2$ or WS$_2$ is not very large, it can be of importance
for Se compounds (with a larger atomic mass than sulfur),
as MoSe$_2$ or WSe$_2$.  The role of the SOC on the chalcogen atom
will be analyzed in more detail in Section~\ref{sec:chalsoc}.

Finally, we can note that, as previously discussed
in Ref. \onlinecite{ZCS11} using first principles
calculations,
the SOC induces a finite band splitting in single-layer
systems also at the Q point, with a corresponding spin-polarization.
Also this feature is nicely captured by our tight-binding model
in the presence of atomic SOC on both chalcogen and transition metal
atoms, as shown in Fig. \ref{Fig:FS}(c) where we plot the Fermi surfaces
of an electron-doped system with a Fermi level cutting only
the lower conduction band at Q.
As we can see, the TB model is able not only to reproduce
the band splitting, but also to point out a strong degree of
entanglement in this point of the BZ,
with Fermi pockets with a strong spin polarization,
and with an alternating polarization of the entangled
spin/valley/orbital degrees of freedom along the six
inequivalent valleys.\cite{YC14}
On the microscopic ground, we can notice that
the main orbital character of the conduction bands at the Q point
is due to a roughly equal distribution of the $d_{x^2-y^2}$ and $d_{xy}$
orbitals of the transition metal $M$, and of the
$p_x$ and $p_y$ orbitals of the chalcogen atom $X$.
Given the presence of a large contribution from both $p$- and
$d$-orbitals, we expect these states to stem from a  strong
hybridization between $X$ and $M$ atoms, and hence to be highly sensitive
to uniform and local strains and lattice distortions.\cite{CS13}
In addition, it should be kept in mind that the minimum of the conduction
band at Q becomes the effective band edge in bilayer and multilayer compounds
(as well as in strained single-layer systems).
These considerations thus suggest  the minima of the conduction band
at the Q point as the most promising states for tuning
the spin/orbital/valley entanglement in these materials
by means of strain engineering\cite{CS13} or (in multilayer systems)
by means of electric fields.\cite{WX13}

\subsection{Effects of the chalcogen atom SOC on the band structure.}
\label{sec:chalsoc}

\begin{figure}[t]
\begin{center}
\mbox{
\includegraphics[width=1\columnwidth]{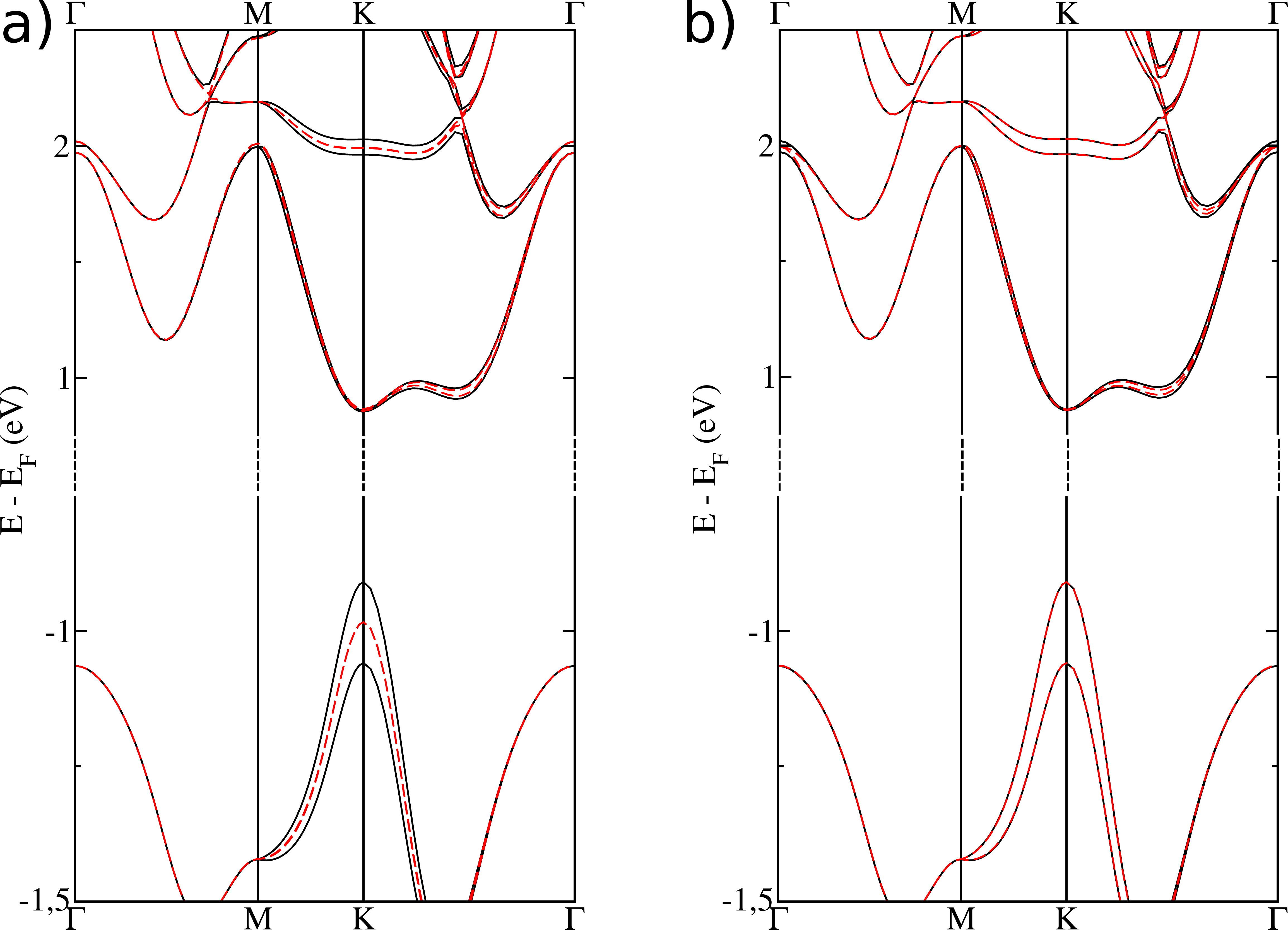}
} 
\end{center}
\caption{Tight-binding band structure of MoS$_2$ including SOC. Solid black lines corresponds to the TB bands using $\lambda_{\rm Mo}$ and $\lambda_{\rm S}$ as given in Table \ref{Tab:TB}. Red dashed lines in (a) corresponds to $\lS=0.052$~eV and $\lMo=0$. Red dashed lines in (b) correspond to $\lMo=0.075$~eV and $\lS=0$.  }
\label{Fig:LambdaMovsS}
\end{figure}

Most of the existing theoretical works have focused on the effects
of the spin-orbit interaction associated with the transition metal atom.
Less attention has been paid, in general, to the SOC induced
by the chalcogen atom.
As we have seen in the previous section, however,
the role of the SOC can be remarkably relevant also at the Q point
of the BZ, resulting in a strong spin/orbital/valley entanglement
also in this point, with the advantage to be extremely sensitive
to the $M$-$X$ hybridization and hence to the lattice effects.
In addition, since the orbital content in this point
is a mixture of $d$ and $p$ orbitals of the metal and the chalcogen atoms,
the spin-orbit coupling is expected to be significantly driven
not only by the $d$-orbital of the transition metal $M$, but also by
the $p$ orbitals of the chalcogen $X$ atom. 
Tight-binding models can be quite useful to investigate this issue
since we can easily tune the atomic SOC,
keeping all the remaining Hamiltonian (Slater-Koster) parameters fixed,
which permits to isolate the effects of the modified SOC
without involving other structural and electronic changes.
Figure~\ref{Fig:LambdaMovsS} shows the  effect of  removing the SOC 
on either the Mo or the S atoms for the case of the single-layer of MoS$_2$. 
While the splitting of the valence band at the K point is
fully caused by the SOC on the transition metal, the 
contributions to the splitting of the conduction band at Q 
from Mo and S are comparable.

\begin{figure*}
   \includegraphics[scale=1]{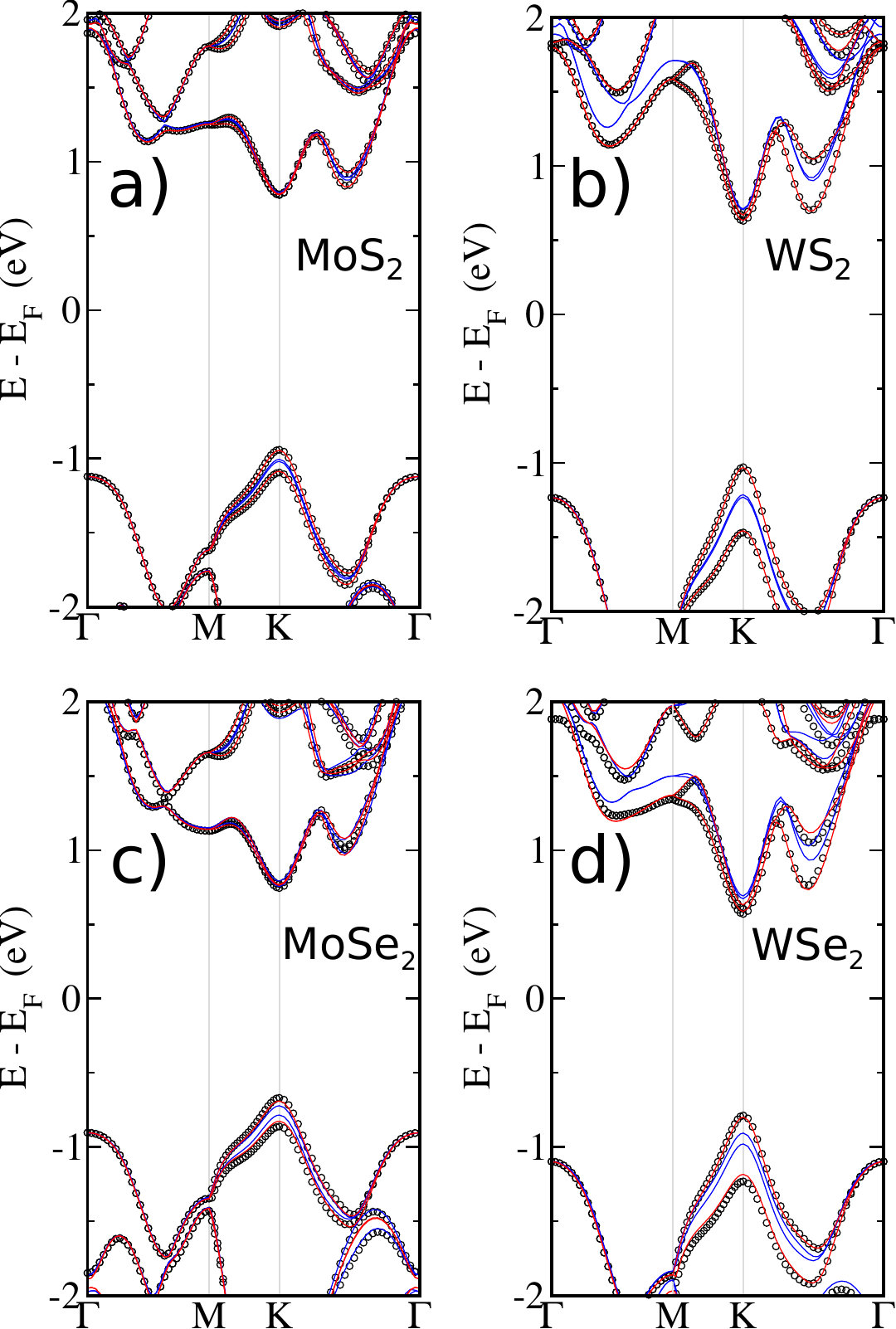}
   \caption{DFT band structure of the four compounds MoS$_2$, WS$_2$, MoSe$_2$ and WSe$_2$. Black circles show the results when the SOC on all the atoms are included. Red (blue) lines correspond to the removal of the SOC on the chalcogen (transition metal) atoms.}     
   \label{fig:SO_NoSO_DFT}
\end{figure*}

We can validate these findings by performing DFT calculations on the four compounds $MX_2$
with $M=$ Mo, W and $X=$ S, Se (all of them done using the experimental structure).  
In the DFT calculation, we can also turn on and off the SOC 
on a particular species, by removing the SO component of the pseudopotential.\cite{kleinman_1980} 
Fig. \ref{fig:SO_NoSO_DFT} shows the DFT results for the four compounds, including the SOC on all the atoms,
and removing this coupling on either the chalcogen or the transition metal. In particular, the DFT results for MoS$_2$ shown in Fig. \ref{fig:SO_NoSO_DFT}(a) agrees reasonably well with those of Fig. \ref{Fig:LambdaMovsS}, signaling that the SOC splitting of the bands in the sulfur compounds is dominated by the contribution due to the transition metal atom.

The importance of the SOC of the chalcogen atom is expected to
be more remarkable for heavier atoms,
such as selenium, instead of sulfur.
In Fig. \ref{fig:SO_NoSO_DFT}(c) and (d) we show the DFT band structure for MoSe$_2$ and WSe$_2$, isolating the contribution of the SOC due to the metal and to the chalcogen atoms. As expected, we observe that a relevant contribution to the SOC splitting of the bands is due to the Se atom. This can be seen by a noticeable splitting of the blue lines in Fig. \ref{fig:SO_NoSO_DFT}(c) and (d) (for which the SOC due to the metal $M$ has been switched off) which is governed by the spin-orbit interaction of the Se atoms. Interestingly, this effect is not relevant only at the Q point of the conduction band, but also at the K point of the valence band, for which the orbital weight of the $p_x$ and $p_y$ orbitals of Se is of {\it only} $\sim 20\%$.\cite{CG13}  We conclude that, although for the MoS$_2$ and WS$_2$ the effect of the SOC of the chalcogenides does not have much effect on the band structure, when S is changed by Se, the effect is much noticeable.

\subsection{Spin-Valley-Layer coupling in Bilayer {\em MX}$_2$}

\begin{figure}[t]
\begin{center}
\mbox{
\includegraphics[width=1.\columnwidth]{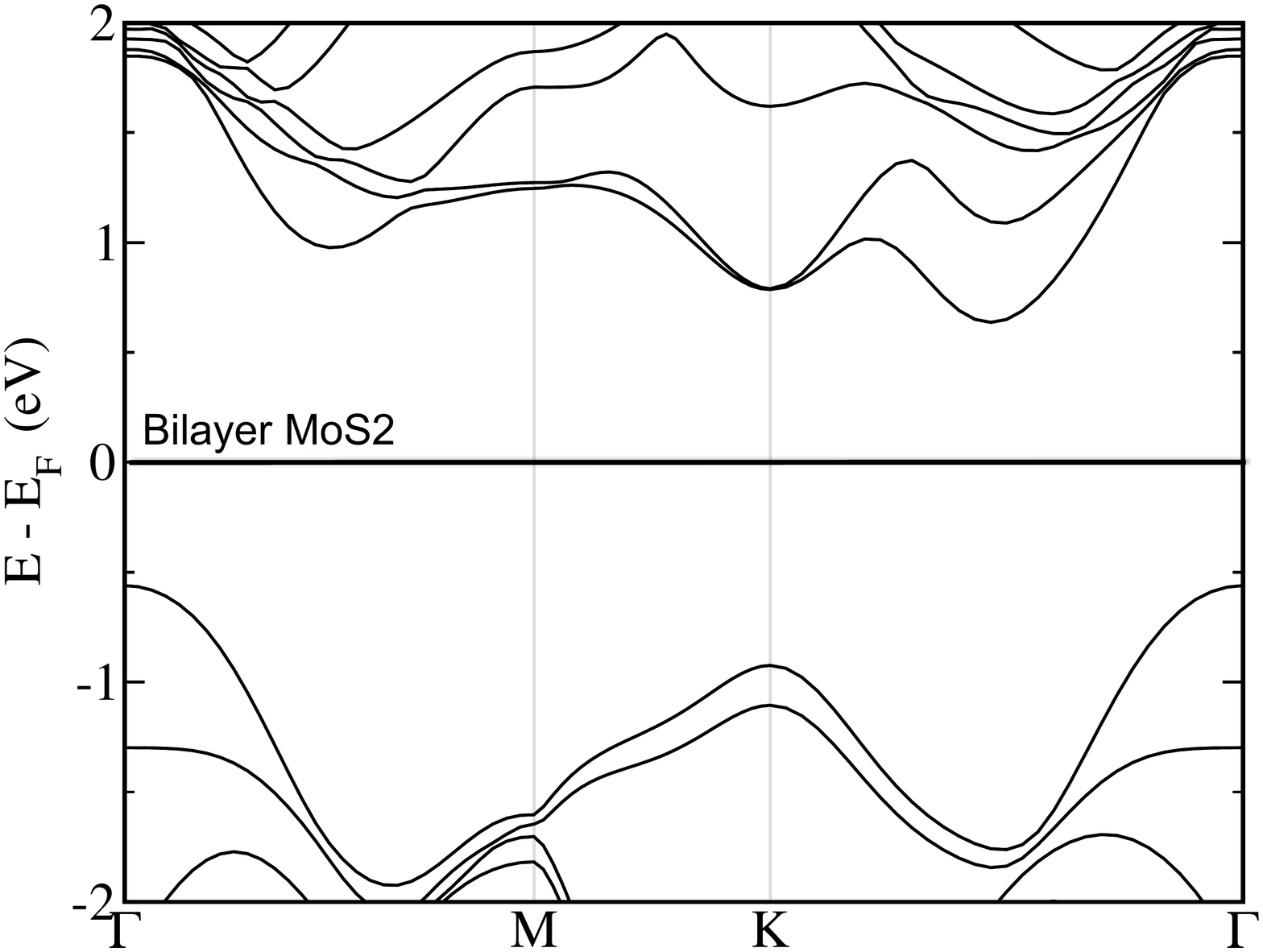}
} 
\end{center}
\begin{center}
\mbox{
\includegraphics[width=1.\columnwidth]{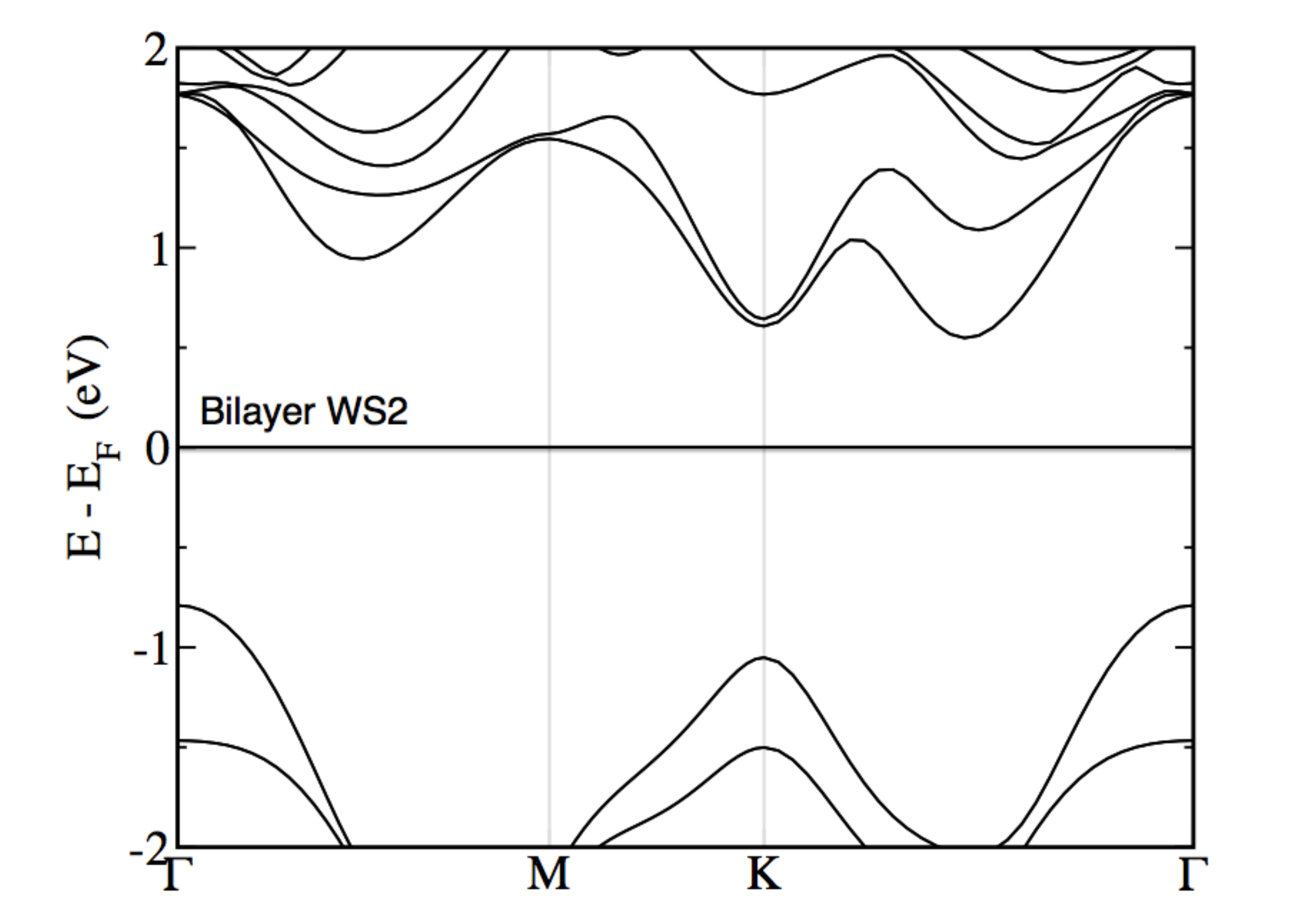}
} 
\end{center}
\caption{Band structure of bilayer MoS$_2$ and WS$_2$
obtained from DFT calculations.
The combined effect of inter-layer hopping and spin-orbit interaction
drives the minimum of the conduction band to the Q point,
and the maximum of the valence band to the $\Gamma$ point (see text) .
}
\label{Fig:BandsBilayerWS2}
\end{figure}

Of special interest is the case of bilayer TMD,
corresponding to a stack of two single layers in-plane rotated by $180^\circ$ 
with respect to each other, such that the transition metal atoms of one layer are above the chalcogen atoms of the other layer. The two layers are bound by means of weak van der Waals
interactions.
The inter-layer hopping of electrons between different layers leads
to a strong modification of the band structure, driving a transition
from a direct gap semiconductor in single-layer systems
to an indirect gap semiconductor in bilayer and multi-layer compounds.
The inter-layer hopping
links mainly the $p$ orbitals of the chalcogen atoms $X$ of different
layers.\cite{CG13}
The result of this hopping is a splitting of the maximum of the
valence band at the $\Gamma$ point, which becomes the effective
valence band edge, as well as a splitting of the minimum of the
conduction band at the Q point which becomes the absolute minimum
of the conduction band.
This situation
is shown in Fig. \ref{Fig:BandsBilayerWS2}, where we report
the band structure of bilayer MoS$_2$ and WS$_2$ calculated by DFT methods.
A qualitative similar feature is observed also
in other bilayer compounds,  as MoSe$_2$ or WSe$_2$.

Contrary to single-layer {\em MX}$_2$, bilayer {\em MX}$_2$ presents
point-center inversion symmetry.\cite{WX13,GY13,ZC13}
Therefore, as we have discussed for the bulk case, the corresponding band structure
remains spin degenerate even in the presence of SOC.
However, since the SOC Hamiltonian does not couple orbitals of
different layers, each single band preserves a finite entanglement between
spin, valley and the layer index.
Such spin-valley-layer coupling has been discussed in
Ref. \onlinecite{GY13},
where the authors focused on the relevance of this effect at the K
point of the valence band.
Here we notice that the same effect occurs also for the conduction
band, and it can be thus relevant for electron-doped samples.
Indeed
for slightly electron-doped bilayer MoS$_2$ and WS$_2$
the Fermi surface
presents six pockets centered at the inequivalent Q valleys of the BZ,
and no pockets at the K and K' valleys.
Interestingly, the SOC for the TMD families with
stronger spin-orbit interaction, like WS$_2$ and WSe$_2$, can be larger than the
inter-layer hopping, enhancing the spin/layer/valley entanglement.
Then, although inversion symmetry forces each Fermi pocket to be spin
degenerate,
the layer polarization makes that each layer
contributes with opposite spin in alternating valleys.
This property can be of interest for {\it valleytronics} devices: by partially filling only one of the two subbands at the Q point of the conduction band, one would have a situation in which the upper layer contributes to three of the six valleys with spin-$\uparrow$, and with spin-$\downarrow$ to the other three valleys, whereas the opposite contribution is inferred from the bottom layer. This spin-valley coupling scenario resembles that of single-layer and bilayer {\em MX}$_2$ discussed in the literature, but for electron-doped samples, which is the kind of doping most commonly reported for those materials. Although we have focused in this section in the most simple multi-layer compound, which is the bilayer $MX_2$, the physics discussed above applies also to any multi-layer TMD with an even number of layers, because they contain the same symmetry properties as that of bilayer $MX_2$ discussed here.

\section{Conclusions}

In conclusion, we have studied the effect of SOC in the band structure of TMD. We have used a tight-binding model, valid for single-layer samples as well as for multi-layer samples, which includes the SO interaction of both, the transition metal and the chalcogen atoms. The band structure obtained from the TB model has been compared to DFT calculations for MoS$_2$ and WS$_2$. Based on the orbital character at each relevant point of the Brillouin zone, we have discussed the origin and main features of the SOC effects at the different band edges. In particular we have found that, for the cases of interest here, spin-flip processes are negligible in the SOC Hamiltonian. This allows to highly simplify the model, making possible to construct a reduced TB Hamiltonian which contains the orbital character and SOC which is relevant for the description of the system around the gap. Special attention has been paid to the role of the SOC associated to the chalcogen atom. In fact, whereas most of the previous works has focused on the SOC associated to the metal atom (which is indeed the responsible for the large splitting of the valence band at the K point) here we have shown that the SOC associated to the chalcogen atom may be important at the Q point of the conduction band, and at the K point of the valence band, especially for MoSe$_2$ and WSe$_2$. Finally, we have considered the effect of SOC in bilayer TMD. Whereas for single-layer {\em MX}$_2$ inversion asymmetry leads to spin-valley coupling, the band edges of bilayer TMD are spin degenerate. However, since inter-layer hopping conserves the spin, the spin physics can be exploited in bilayer {\em MX}$_2$ due to spin-valley-layer coupling. Whereas this issue has been recently studied in detail for hole-doped samples,\cite{GY13} here we have argued that a similar effect can be expected for slightly electron-doped samples.

\acknowledgments

We thank H. Ochoa and S. Gallego for useful discussions. R.R., M.P.L.-S. and F.G. acknowledge financial support from MINECO, Spain,
through grant FIS2011-23713, and the European Union,
through grant 290846. R. R. acknowledges financial support
from the Juan de la Cierva Program (MINECO, Spain).
E.C. acknowledges support from the
European project FP7-PEOPLE-2013-CIG ``LSIE\_2D''
and Italian National Miur Prin project 20105ZZTSE. 
J.A.S.-G. and P.O. ackowledge support from Spanish MINECO
(Grants No. FIS2012-37549-C05-02 with joint financing by FEDER Funds from the European Union, and
No. CSD2007-00050). J.A.S.-G. was supported by an FPI Fellowship from MINECO. The authors thankfully acknowledges the computer resources, technical expertise and assistance provided by the Red Espa\~nola de Supercomputaci\'on.

\appendix

%%%%%%%%%%%%%%%%%%%%%%%%%%%%%%%%%%%%%%%%%%%%%%%%%%%%%%%%%%%%
%%%%%%%%%%%%%%%%%%%%%%%%%%%%%%%%%%%%%%%%%%%%%%%%%%%%%%%%%%%%

\section{SOC Hamiltonian}\label{App:SOC}

In this appendix we provide the explicit expression
of the matrices $\hat{M}_{\rm EE}^{\sigma\sigma}$,
$\hat{M}_{\rm OO}^{\sigma\sigma}$,
$\hat{M}_{\rm EO}^{\sigma\bar{\sigma}}$,
$\hat{M}_{\rm OE}^{\sigma\bar{\sigma}}$,
describing the local atomic spin-orbit interaction
on both $M$ and $X$ atoms. We have:
\begin{eqnarray}
\hat{M}_{\rm EE}^{\uparrow\uparrow}
&=&
\left(
\begin{array}{cccccc}
0 &0 &0 &0 &0 &0 \\
 0&0 & -i\lambda_{M} &0  &0  &0  \\
 0 & i\lambda_{M} & 0 & 0 & 0 &0 \\
 0& 0 &0 & 0 &  -i\lambda_{X}/2 & 0\\
0 &0 & 0 & i\lambda_{X}/2 & 0 & 0\\
0& 0&0& 0 &0 &0
\end{array}
\right),
\label{Eq:M11}
\end{eqnarray}
\begin{eqnarray}
\hat{M}_{\rm EE}^{\downarrow\downarrow}
&=&
-\hat{M}_{\rm EE}^{\uparrow\uparrow},
\label{Eq:M22}
\end{eqnarray}
\begin{eqnarray}
\hat{M}_{\rm OO}^{\uparrow\uparrow}
&=&
\frac{1}{2}
\left(
\begin{array}{ccccc}
0 & -i\lambda_{M} & 0 & 0 & 0\\
i\lambda_{M} & 0 & 0 & 0 & 0\\
0 & 0 & 0 & -i\lambda_{X} & 0\\
0 & 0  & i\lambda_{X} & 0 & 0\\
0 & 0  & 0 & 0 & 0
\end{array}
\right),
\label{Eq:M33}
\end{eqnarray}
\begin{eqnarray}
\hat{M}_{\rm OO}^{\downarrow\downarrow}
&=&
-\hat{M}_{\rm OO}^{\uparrow\uparrow},
\label{Eq:M44}
\end{eqnarray}
\begin{eqnarray}
\hat{M}_{\rm EO}^{\uparrow\downarrow}
&=&
\frac{1}{2}
\left(
\begin{array}{ccccc}
-\sqrt{3}\lambda_{M} &i\sqrt{3}\lambda_{M} &0 &0 &0  \\
 \lambda_{M}& i\lambda_{M} &0  &0 & 0     \\
 -i\lambda_{M} & \lambda_{M} & 0 & 0 & 0  \\
 0& 0 &0 & 0 &  \lambda_{X} \\
0 &0 & 0 & 0 &-i\lambda_{X}  \\
0& 0&-\lambda_{X}& i\lambda_{X} &0 
\end{array}
\right),
\label{Eq:M12}
\end{eqnarray}
\begin{eqnarray}
\hat{M}_{\rm OE}^{\downarrow\uparrow}
&=&
\left(\hat{M}_{\rm EO}^{\uparrow\downarrow}\right)^{\dagger},
\label{Eq:M41}
\end{eqnarray}
\begin{eqnarray}
\hat{M}_{\rm EO}^{\downarrow\uparrow}
&=&
\frac{1}{2}
\left(
\begin{array}{ccccc}
\sqrt{3}\lambda_{M} &i\sqrt{3}\lambda_{M} &0 &0 &0  \\
 -\lambda_{M}& i\lambda_{M} &0  &0 & 0     \\
 -i\lambda_{M} & -\lambda_{M} & 0 & 0 & 0  \\
 0& 0 &0 & 0 &  -\lambda_{X} \\
0 &0 & 0 & 0 &-i\lambda_{X}  \\
0& 0&\lambda_{X}& i\lambda_{X} &0 
\end{array}
\right)
\label{Eq:M23}
\end{eqnarray}
and%
\begin{eqnarray}
\hat{M}_{\rm OE}^{\uparrow\downarrow}
&=&
\left(\hat{M}_{\rm EO}^{\downarrow\uparrow}\right)^{\dagger},
\label{Eq:M32}
\end{eqnarray}
In the above matrices we have used the short notation
$\lambda_{M}$ for the SOC of the metal (Mo or W)
and $\lambda_X$ for the SOC of the chalcogen (S or Se).

\bibliography{BibliogrGrafeno}

\newcommand{\npb}{Nucl. Phys.}\newcommand{\adv}{Adv.
  Phys.}\newcommand{\epl}{Europhys. Lett.}
\begin{thebibliography}{49}
\expandafter\ifx\csname natexlab\endcsname\relax\def\natexlab#1{#1}\fi
\expandafter\ifx\csname bibnamefont\endcsname\relax
  \def\bibnamefont#1{#1}\fi
\expandafter\ifx\csname bibfnamefont\endcsname\relax
  \def\bibfnamefont#1{#1}\fi
\expandafter\ifx\csname citenamefont\endcsname\relax
  \def\citenamefont#1{#1}\fi
\expandafter\ifx\csname url\endcsname\relax
  \def\url#1{\texttt{#1}}\fi
\expandafter\ifx\csname urlprefix\endcsname\relax\def\urlprefix{URL }\fi
\providecommand{\bibinfo}[2]{#2}
\providecommand{\eprint}[2][]{\url{#2}}

\bibitem[{\citenamefont{Wang et~al.}(2012)\citenamefont{Wang, Kalantar-Zadeh,
  Kis, Coleman, and Strano}}]{WS12}
\bibinfo{author}{\bibfnamefont{Q.~H.} \bibnamefont{Wang}},
  \bibinfo{author}{\bibfnamefont{K.}~\bibnamefont{Kalantar-Zadeh}},
  \bibinfo{author}{\bibfnamefont{A.}~\bibnamefont{Kis}},
  \bibinfo{author}{\bibfnamefont{J.~N.} \bibnamefont{Coleman}},
  \bibnamefont{and} \bibinfo{author}{\bibfnamefont{M.~S.}
  \bibnamefont{Strano}}, \bibinfo{journal}{Nature Nanotech.}
  \textbf{\bibinfo{volume}{7}}, \bibinfo{pages}{699} (\bibinfo{year}{2012}).

\bibitem[{\citenamefont{Mak et~al.}(2010)\citenamefont{Mak, Lee, Hone, Shan,
  and Heinz}}]{MH10}
\bibinfo{author}{\bibfnamefont{K.~F.} \bibnamefont{Mak}},
  \bibinfo{author}{\bibfnamefont{C.}~\bibnamefont{Lee}},
  \bibinfo{author}{\bibfnamefont{J.}~\bibnamefont{Hone}},
  \bibinfo{author}{\bibfnamefont{J.}~\bibnamefont{Shan}}, \bibnamefont{and}
  \bibinfo{author}{\bibfnamefont{T.~F.} \bibnamefont{Heinz}},
  \bibinfo{journal}{Phys. Rev. Lett.} \textbf{\bibinfo{volume}{105}},
  \bibinfo{pages}{136805} (\bibinfo{year}{2010}).

\bibitem[{\citenamefont{Cappelluti et~al.}(2013)\citenamefont{Cappelluti,
  Rold\'an, Silva-Guill\'en, Ordej\'on, and Guinea}}]{CG13}
\bibinfo{author}{\bibfnamefont{E.}~\bibnamefont{Cappelluti}},
  \bibinfo{author}{\bibfnamefont{R.}~\bibnamefont{Rold\'an}},
  \bibinfo{author}{\bibfnamefont{J.~A.} \bibnamefont{Silva-Guill\'en}},
  \bibinfo{author}{\bibfnamefont{P.}~\bibnamefont{Ordej\'on}},
  \bibnamefont{and} \bibinfo{author}{\bibfnamefont{F.}~\bibnamefont{Guinea}},
  \bibinfo{journal}{Phys. Rev. B} \textbf{\bibinfo{volume}{88}},
  \bibinfo{pages}{075409} (\bibinfo{year}{2013}).

\bibitem[{\citenamefont{Zhu et~al.}(2011)\citenamefont{Zhu, Cheng, and
  Schwingenschl\"ogl}}]{ZCS11}
\bibinfo{author}{\bibfnamefont{Z.~Y.} \bibnamefont{Zhu}},
  \bibinfo{author}{\bibfnamefont{Y.~C.} \bibnamefont{Cheng}}, \bibnamefont{and}
  \bibinfo{author}{\bibfnamefont{U.}~\bibnamefont{Schwingenschl\"ogl}},
  \bibinfo{journal}{Phys. Rev. B} \textbf{\bibinfo{volume}{84}},
  \bibinfo{pages}{153402} (\bibinfo{year}{2011}).

\bibitem[{\citenamefont{Xiao et~al.}(2012)\citenamefont{Xiao, Liu, Feng, Xu,
  and Yao}}]{XY12}
\bibinfo{author}{\bibfnamefont{D.}~\bibnamefont{Xiao}},
  \bibinfo{author}{\bibfnamefont{G.-B.} \bibnamefont{Liu}},
  \bibinfo{author}{\bibfnamefont{W.}~\bibnamefont{Feng}},
  \bibinfo{author}{\bibfnamefont{X.}~\bibnamefont{Xu}}, \bibnamefont{and}
  \bibinfo{author}{\bibfnamefont{W.}~\bibnamefont{Yao}},
  \bibinfo{journal}{Phys. Rev. Lett.} \textbf{\bibinfo{volume}{108}},
  \bibinfo{pages}{196802} (\bibinfo{year}{2012}).

\bibitem[{\citenamefont{Feng et~al.}(2012)\citenamefont{Feng, Yao, Zhu, Zhou,
  Yao, and Xiao}}]{FX12}
\bibinfo{author}{\bibfnamefont{W.}~\bibnamefont{Feng}},
  \bibinfo{author}{\bibfnamefont{Y.}~\bibnamefont{Yao}},
  \bibinfo{author}{\bibfnamefont{W.}~\bibnamefont{Zhu}},
  \bibinfo{author}{\bibfnamefont{J.}~\bibnamefont{Zhou}},
  \bibinfo{author}{\bibfnamefont{W.}~\bibnamefont{Yao}}, \bibnamefont{and}
  \bibinfo{author}{\bibfnamefont{D.}~\bibnamefont{Xiao}},
  \bibinfo{journal}{Phys. Rev. B} \textbf{\bibinfo{volume}{86}},
  \bibinfo{pages}{165108} (\bibinfo{year}{2012}).

\bibitem[{\citenamefont{Shan et~al.}(2013)\citenamefont{Shan, Lu, and
  Xiao}}]{SX13}
\bibinfo{author}{\bibfnamefont{W.-Y.} \bibnamefont{Shan}},
  \bibinfo{author}{\bibfnamefont{H.-Z.} \bibnamefont{Lu}}, \bibnamefont{and}
  \bibinfo{author}{\bibfnamefont{D.}~\bibnamefont{Xiao}},
  \bibinfo{journal}{Phys. Rev. B} \textbf{\bibinfo{volume}{88}},
  \bibinfo{pages}{125301} (\bibinfo{year}{2013}).

\bibitem[{\citenamefont{Rostami et~al.}(2013)\citenamefont{Rostami, Moghaddam,
  and Asgari}}]{RMA13}
\bibinfo{author}{\bibfnamefont{H.}~\bibnamefont{Rostami}},
  \bibinfo{author}{\bibfnamefont{A.~G.} \bibnamefont{Moghaddam}},
  \bibnamefont{and} \bibinfo{author}{\bibfnamefont{R.}~\bibnamefont{Asgari}},
  \bibinfo{journal}{Phys. Rev. B} \textbf{\bibinfo{volume}{88}},
  \bibinfo{pages}{085440} (\bibinfo{year}{2013}).

\bibitem[{\citenamefont{Rose et~al.}(2013)\citenamefont{Rose, Goerbig, and
  Pi\'echon}}]{RGP13}
\bibinfo{author}{\bibfnamefont{F.}~\bibnamefont{Rose}},
  \bibinfo{author}{\bibfnamefont{M.~O.} \bibnamefont{Goerbig}},
  \bibnamefont{and}
  \bibinfo{author}{\bibfnamefont{F.}~\bibnamefont{Pi\'echon}},
  \bibinfo{journal}{Phys. Rev. B} \textbf{\bibinfo{volume}{88}},
  \bibinfo{pages}{125438} (\bibinfo{year}{2013}).

\bibitem[{\citenamefont{Liu et~al.}(2013)\citenamefont{Liu, Shan, Yao, Yao, and
  Xiao}}]{LX13}
\bibinfo{author}{\bibfnamefont{G.-B.} \bibnamefont{Liu}},
  \bibinfo{author}{\bibfnamefont{W.-Y.} \bibnamefont{Shan}},
  \bibinfo{author}{\bibfnamefont{Y.}~\bibnamefont{Yao}},
  \bibinfo{author}{\bibfnamefont{W.}~\bibnamefont{Yao}}, \bibnamefont{and}
  \bibinfo{author}{\bibfnamefont{D.}~\bibnamefont{Xiao}},
  \bibinfo{journal}{Phys. Rev. B} \textbf{\bibinfo{volume}{88}},
  \bibinfo{pages}{085433} (\bibinfo{year}{2013}).

\bibitem[{\citenamefont{Cao et~al.}(2012)\citenamefont{Cao, Wang, Han, Ye, Zhu,
  Shi, Niu, Tan, Wang, Liu et~al.}}]{Cao_etal_2012}
\bibinfo{author}{\bibfnamefont{T.}~\bibnamefont{Cao}},
  \bibinfo{author}{\bibfnamefont{G.}~\bibnamefont{Wang}},
  \bibinfo{author}{\bibfnamefont{W.}~\bibnamefont{Han}},
  \bibinfo{author}{\bibfnamefont{H.}~\bibnamefont{Ye}},
  \bibinfo{author}{\bibfnamefont{C.}~\bibnamefont{Zhu}},
  \bibinfo{author}{\bibfnamefont{J.}~\bibnamefont{Shi}},
  \bibinfo{author}{\bibfnamefont{Q.}~\bibnamefont{Niu}},
  \bibinfo{author}{\bibfnamefont{P.}~\bibnamefont{Tan}},
  \bibinfo{author}{\bibfnamefont{E.}~\bibnamefont{Wang}},
  \bibinfo{author}{\bibfnamefont{B.}~\bibnamefont{Liu}}, \bibnamefont{et~al.},
  \bibinfo{journal}{Nature Commun.} \textbf{\bibinfo{volume}{3}},
  \bibinfo{pages}{887} (\bibinfo{year}{2012}).

\bibitem[{\citenamefont{Zeng et~al.}(2012)\citenamefont{Zeng, Dai, Yao, Xiao,
  and Cui}}]{Zeng_etal_2012}
\bibinfo{author}{\bibfnamefont{H.}~\bibnamefont{Zeng}},
  \bibinfo{author}{\bibfnamefont{J.}~\bibnamefont{Dai}},
  \bibinfo{author}{\bibfnamefont{W.}~\bibnamefont{Yao}},
  \bibinfo{author}{\bibfnamefont{D.}~\bibnamefont{Xiao}}, \bibnamefont{and}
  \bibinfo{author}{\bibfnamefont{X.}~\bibnamefont{Cui}},
  \bibinfo{journal}{Nature Nanotech.} \textbf{\bibinfo{volume}{7}},
  \bibinfo{pages}{490} (\bibinfo{year}{2012}).

\bibitem[{\citenamefont{Mak et~al.}(2012)\citenamefont{Mak, He, Sahn, and
  Heinz}}]{Mak_etal_2012}
\bibinfo{author}{\bibfnamefont{K.~F.} \bibnamefont{Mak}},
  \bibinfo{author}{\bibfnamefont{K.}~\bibnamefont{He}},
  \bibinfo{author}{\bibfnamefont{J.}~\bibnamefont{Sahn}}, \bibnamefont{and}
  \bibinfo{author}{\bibfnamefont{T.~F.} \bibnamefont{Heinz}},
  \bibinfo{journal}{Nature Nanotech.} \textbf{\bibinfo{volume}{7}},
  \bibinfo{pages}{494} (\bibinfo{year}{2012}).

\bibitem[{\citenamefont{Wu et~al.}(2013)\citenamefont{Wu, Ross, Liu, Aivazian,
  Jones, Fei, Zhu, Xiao, Yao, Cobden et~al.}}]{WX13}
\bibinfo{author}{\bibfnamefont{S.}~\bibnamefont{Wu}},
  \bibinfo{author}{\bibfnamefont{J.~S.} \bibnamefont{Ross}},
  \bibinfo{author}{\bibfnamefont{G.-B.} \bibnamefont{Liu}},
  \bibinfo{author}{\bibfnamefont{G.}~\bibnamefont{Aivazian}},
  \bibinfo{author}{\bibfnamefont{A.}~\bibnamefont{Jones}},
  \bibinfo{author}{\bibfnamefont{Z.}~\bibnamefont{Fei}},
  \bibinfo{author}{\bibfnamefont{W.}~\bibnamefont{Zhu}},
  \bibinfo{author}{\bibfnamefont{D.}~\bibnamefont{Xiao}},
  \bibinfo{author}{\bibfnamefont{W.}~\bibnamefont{Yao}},
  \bibinfo{author}{\bibfnamefont{D.}~\bibnamefont{Cobden}},
  \bibnamefont{et~al.}, \bibinfo{journal}{Nature Physics}
  \textbf{\bibinfo{volume}{9}}, \bibinfo{pages}{149} (\bibinfo{year}{2013}).

\bibitem[{\citenamefont{Wang et~al.}(2013)\citenamefont{Wang, Ge, Li, Qiu, Ji,
  Feng, and Sun}}]{WS13}
\bibinfo{author}{\bibfnamefont{Q.}~\bibnamefont{Wang}},
  \bibinfo{author}{\bibfnamefont{S.}~\bibnamefont{Ge}},
  \bibinfo{author}{\bibfnamefont{X.}~\bibnamefont{Li}},
  \bibinfo{author}{\bibfnamefont{J.}~\bibnamefont{Qiu}},
  \bibinfo{author}{\bibfnamefont{Y.}~\bibnamefont{Ji}},
  \bibinfo{author}{\bibfnamefont{J.}~\bibnamefont{Feng}}, \bibnamefont{and}
  \bibinfo{author}{\bibfnamefont{D.}~\bibnamefont{Sun}}, \bibinfo{journal}{ACS
  Nano} \textbf{\bibinfo{volume}{7}}, \bibinfo{pages}{11087}
  (\bibinfo{year}{2013}).

\bibitem[{\citenamefont{{Zeng} et~al.}(2013)\citenamefont{{Zeng}, {Liu}, {Dai},
  {Yan}, {Zhu}, {He}, {Xie}, {Xu}, {Chen}, {Yao} et~al.}}]{ZC13}
\bibinfo{author}{\bibfnamefont{H.}~\bibnamefont{{Zeng}}},
  \bibinfo{author}{\bibfnamefont{G.-B.} \bibnamefont{{Liu}}},
  \bibinfo{author}{\bibfnamefont{J.}~\bibnamefont{{Dai}}},
  \bibinfo{author}{\bibfnamefont{Y.}~\bibnamefont{{Yan}}},
  \bibinfo{author}{\bibfnamefont{B.}~\bibnamefont{{Zhu}}},
  \bibinfo{author}{\bibfnamefont{R.}~\bibnamefont{{He}}},
  \bibinfo{author}{\bibfnamefont{L.}~\bibnamefont{{Xie}}},
  \bibinfo{author}{\bibfnamefont{S.}~\bibnamefont{{Xu}}},
  \bibinfo{author}{\bibfnamefont{X.}~\bibnamefont{{Chen}}},
  \bibinfo{author}{\bibfnamefont{W.}~\bibnamefont{{Yao}}},
  \bibnamefont{et~al.}, \bibinfo{journal}{Scientific Reports}
  \textbf{\bibinfo{volume}{3}}, \bibinfo{eid}{1608} (\bibinfo{year}{2013}).

\bibitem[{\citenamefont{Ochoa and Rold\'an}(2013)}]{OR13}
\bibinfo{author}{\bibfnamefont{H.}~\bibnamefont{Ochoa}} \bibnamefont{and}
  \bibinfo{author}{\bibfnamefont{R.}~\bibnamefont{Rold\'an}},
  \bibinfo{journal}{Phys. Rev. B} \textbf{\bibinfo{volume}{87}},
  \bibinfo{pages}{245421} (\bibinfo{year}{2013}).

\bibitem[{\citenamefont{Cheiwchanchamnangij and Lambrecht}(2012)}]{CL12}
\bibinfo{author}{\bibfnamefont{T.}~\bibnamefont{Cheiwchanchamnangij}}
  \bibnamefont{and} \bibinfo{author}{\bibfnamefont{W.~R.~L.}
  \bibnamefont{Lambrecht}}, \bibinfo{journal}{Phys. Rev. B}
  \textbf{\bibinfo{volume}{85}}, \bibinfo{pages}{205302}
  (\bibinfo{year}{2012}).

\bibitem[{\citenamefont{Ko\ifmmode~\acute{s}\else \'{s}\fi{}mider and
  Fern\'andez-Rossier}(2013)}]{Kosmider_etal_2013}
\bibinfo{author}{\bibfnamefont{K.}~\bibnamefont{Ko\ifmmode~\acute{s}\else
  \'{s}\fi{}mider}} \bibnamefont{and}
  \bibinfo{author}{\bibfnamefont{J.}~\bibnamefont{Fern\'andez-Rossier}},
  \bibinfo{journal}{Phys. Rev. B} \textbf{\bibinfo{volume}{87}},
  \bibinfo{pages}{075451} (\bibinfo{year}{2013}).

\bibitem[{\citenamefont{Molina-S\'anchez
  et~al.}(2013)\citenamefont{Molina-S\'anchez, Sangalli, Hummer, Marini, and
  Wirtz}}]{MW13}
\bibinfo{author}{\bibfnamefont{A.}~\bibnamefont{Molina-S\'anchez}},
  \bibinfo{author}{\bibfnamefont{D.}~\bibnamefont{Sangalli}},
  \bibinfo{author}{\bibfnamefont{K.}~\bibnamefont{Hummer}},
  \bibinfo{author}{\bibfnamefont{A.}~\bibnamefont{Marini}}, \bibnamefont{and}
  \bibinfo{author}{\bibfnamefont{L.}~\bibnamefont{Wirtz}},
  \bibinfo{journal}{Phys. Rev. B} \textbf{\bibinfo{volume}{88}},
  \bibinfo{pages}{045412} (\bibinfo{year}{2013}).

\bibitem[{\citenamefont{Song and Dery}(2013)}]{SD13}
\bibinfo{author}{\bibfnamefont{Y.}~\bibnamefont{Song}} \bibnamefont{and}
  \bibinfo{author}{\bibfnamefont{H.}~\bibnamefont{Dery}},
  \bibinfo{journal}{Phys. Rev. Lett.} \textbf{\bibinfo{volume}{111}},
  \bibinfo{pages}{026601} (\bibinfo{year}{2013}).

\bibitem[{\citenamefont{Korm\'anyos et~al.}(2013)\citenamefont{Korm\'anyos,
  Z\'olyomi, Drummond, Rakyta, Burkard, and Fal'ko}}]{KF13}
\bibinfo{author}{\bibfnamefont{A.}~\bibnamefont{Korm\'anyos}},
  \bibinfo{author}{\bibfnamefont{V.}~\bibnamefont{Z\'olyomi}},
  \bibinfo{author}{\bibfnamefont{N.~D.} \bibnamefont{Drummond}},
  \bibinfo{author}{\bibfnamefont{P.}~\bibnamefont{Rakyta}},
  \bibinfo{author}{\bibfnamefont{G.}~\bibnamefont{Burkard}}, \bibnamefont{and}
  \bibinfo{author}{\bibfnamefont{V.~I.} \bibnamefont{Fal'ko}},
  \bibinfo{journal}{Phys. Rev. B} \textbf{\bibinfo{volume}{88}},
  \bibinfo{pages}{045416} (\bibinfo{year}{2013}).

\bibitem[{\citenamefont{Korm\'anyos et~al.}(2014)\citenamefont{Korm\'anyos,
  Z\'olyomi, Drummond, and Burkard}}]{KB13}
\bibinfo{author}{\bibfnamefont{A.}~\bibnamefont{Korm\'anyos}},
  \bibinfo{author}{\bibfnamefont{V.}~\bibnamefont{Z\'olyomi}},
  \bibinfo{author}{\bibfnamefont{N.~D.} \bibnamefont{Drummond}},
  \bibnamefont{and} \bibinfo{author}{\bibfnamefont{G.}~\bibnamefont{Burkard}},
  \bibinfo{journal}{Phys. Rev. X} \textbf{\bibinfo{volume}{4}},
  \bibinfo{pages}{011034} (\bibinfo{year}{2014}).

\bibitem[{\citenamefont{Ko\ifmmode~\acute{s}\else \'{s}\fi{}mider
  et~al.}(2013)\citenamefont{Ko\ifmmode~\acute{s}\else \'{s}\fi{}mider,
  Gonz\'alez, and Fern\'andez-Rossier}}]{KGF13}
\bibinfo{author}{\bibfnamefont{K.}~\bibnamefont{Ko\ifmmode~\acute{s}\else
  \'{s}\fi{}mider}}, \bibinfo{author}{\bibfnamefont{J.~W.}
  \bibnamefont{Gonz\'alez}}, \bibnamefont{and}
  \bibinfo{author}{\bibfnamefont{J.}~\bibnamefont{Fern\'andez-Rossier}},
  \bibinfo{journal}{Phys. Rev. B} \textbf{\bibinfo{volume}{88}},
  \bibinfo{pages}{245436} (\bibinfo{year}{2013}).

\bibitem[{\citenamefont{{Zhao} et~al.}(2013)\citenamefont{{Zhao}, {Ribeiro},
  {Toh}, {Carvalho}, {Kloc}, {Castro Neto}, and {Eda}}}]{ZE13}
\bibinfo{author}{\bibfnamefont{W.}~\bibnamefont{{Zhao}}},
  \bibinfo{author}{\bibfnamefont{R.~M.} \bibnamefont{{Ribeiro}}},
  \bibinfo{author}{\bibfnamefont{M.}~\bibnamefont{{Toh}}},
  \bibinfo{author}{\bibfnamefont{A.}~\bibnamefont{{Carvalho}}},
  \bibinfo{author}{\bibfnamefont{C.}~\bibnamefont{{Kloc}}},
  \bibinfo{author}{\bibfnamefont{A.~H.} \bibnamefont{{Castro Neto}}},
  \bibnamefont{and} \bibinfo{author}{\bibfnamefont{G.}~\bibnamefont{{Eda}}},
  \bibinfo{journal}{ArXiv e-prints}  (\bibinfo{year}{2013}),
  \eprint{1309.0923}.

\bibitem[{\citenamefont{Jin et~al.}(2013)\citenamefont{Jin, Yeh, Zaki, Zhang,
  Sadowski, Al-Mahboob, van~der Zande, Chenet, Dadap, Herman et~al.}}]{JO13}
\bibinfo{author}{\bibfnamefont{W.}~\bibnamefont{Jin}},
  \bibinfo{author}{\bibfnamefont{P.-C.} \bibnamefont{Yeh}},
  \bibinfo{author}{\bibfnamefont{N.}~\bibnamefont{Zaki}},
  \bibinfo{author}{\bibfnamefont{D.}~\bibnamefont{Zhang}},
  \bibinfo{author}{\bibfnamefont{J.~T.} \bibnamefont{Sadowski}},
  \bibinfo{author}{\bibfnamefont{A.}~\bibnamefont{Al-Mahboob}},
  \bibinfo{author}{\bibfnamefont{A.~M.} \bibnamefont{van~der Zande}},
  \bibinfo{author}{\bibfnamefont{D.~A.} \bibnamefont{Chenet}},
  \bibinfo{author}{\bibfnamefont{J.~I.} \bibnamefont{Dadap}},
  \bibinfo{author}{\bibfnamefont{I.~P.} \bibnamefont{Herman}},
  \bibnamefont{et~al.}, \bibinfo{journal}{Phys. Rev. Lett.}
  \textbf{\bibinfo{volume}{111}}, \bibinfo{pages}{106801}
  (\bibinfo{year}{2013}).

\bibitem[{\citenamefont{{Zhang} et~al.}(2014)\citenamefont{{Zhang}, {Chang},
  {Zhou}, {Cui}, {Yan}, {Liu}, {Schmitt}, {Lee}, {Moore}, {Chen}
  et~al.}}]{ZS14}
\bibinfo{author}{\bibfnamefont{Y.}~\bibnamefont{{Zhang}}},
  \bibinfo{author}{\bibfnamefont{T.-R.} \bibnamefont{{Chang}}},
  \bibinfo{author}{\bibfnamefont{B.}~\bibnamefont{{Zhou}}},
  \bibinfo{author}{\bibfnamefont{Y.-T.} \bibnamefont{{Cui}}},
  \bibinfo{author}{\bibfnamefont{H.}~\bibnamefont{{Yan}}},
  \bibinfo{author}{\bibfnamefont{Z.}~\bibnamefont{{Liu}}},
  \bibinfo{author}{\bibfnamefont{F.}~\bibnamefont{{Schmitt}}},
  \bibinfo{author}{\bibfnamefont{J.}~\bibnamefont{{Lee}}},
  \bibinfo{author}{\bibfnamefont{R.}~\bibnamefont{{Moore}}},
  \bibinfo{author}{\bibfnamefont{Y.}~\bibnamefont{{Chen}}},
  \bibnamefont{et~al.}, \bibinfo{journal}{Nature Nanotechnology}
  \textbf{\bibinfo{volume}{9}}, \bibinfo{pages}{111} (\bibinfo{year}{2014}).

\bibitem[{\citenamefont{Li et~al.}(2013)\citenamefont{Li, Zhang, and
  Niu}}]{LZN13}
\bibinfo{author}{\bibfnamefont{X.}~\bibnamefont{Li}},
  \bibinfo{author}{\bibfnamefont{F.}~\bibnamefont{Zhang}}, \bibnamefont{and}
  \bibinfo{author}{\bibfnamefont{Q.}~\bibnamefont{Niu}},
  \bibinfo{journal}{Phys. Rev. Lett.} \textbf{\bibinfo{volume}{110}},
  \bibinfo{pages}{066803} (\bibinfo{year}{2013}).

\bibitem[{\citenamefont{Klinovaja and Loss}(2013)}]{KL13}
\bibinfo{author}{\bibfnamefont{J.}~\bibnamefont{Klinovaja}} \bibnamefont{and}
  \bibinfo{author}{\bibfnamefont{D.}~\bibnamefont{Loss}},
  \bibinfo{journal}{Phys. Rev. B} \textbf{\bibinfo{volume}{88}},
  \bibinfo{pages}{075404} (\bibinfo{year}{2013}).

\bibitem[{\citenamefont{Wang and Wu}(2014)}]{WW14}
\bibinfo{author}{\bibfnamefont{L.}~\bibnamefont{Wang}} \bibnamefont{and}
  \bibinfo{author}{\bibfnamefont{M.}~\bibnamefont{Wu}},
  \bibinfo{journal}{Physics Letters A} \textbf{\bibinfo{volume}{378}},
  \bibinfo{pages}{1336 } (\bibinfo{year}{2014}).

\bibitem[{\citenamefont{{Cazalilla} et~al.}(2013)\citenamefont{{Cazalilla},
  {Ochoa}, and {Guinea}}}]{COG13}
\bibinfo{author}{\bibfnamefont{M.~A.} \bibnamefont{{Cazalilla}}},
  \bibinfo{author}{\bibfnamefont{H.}~\bibnamefont{{Ochoa}}}, \bibnamefont{and}
  \bibinfo{author}{\bibfnamefont{F.}~\bibnamefont{{Guinea}}},
  \bibinfo{journal}{ArXiv e-prints}  (\bibinfo{year}{2013}),
  \eprint{1311.6650}.

\bibitem[{\citenamefont{Goerbig et~al.}(2014)\citenamefont{Goerbig, Montambaux,
  and PiŽchon}}]{GMP14}
\bibinfo{author}{\bibfnamefont{M.~O.} \bibnamefont{Goerbig}},
  \bibinfo{author}{\bibfnamefont{G.}~\bibnamefont{Montambaux}},
  \bibnamefont{and} \bibinfo{author}{\bibfnamefont{F.}~\bibnamefont{PiŽchon}},
  \bibinfo{journal}{EPL (Europhysics Letters)} \textbf{\bibinfo{volume}{105}},
  \bibinfo{pages}{57005} (\bibinfo{year}{2014}).

\bibitem[{\citenamefont{Gallego and Munoz}(1999)}]{GM99}
\bibinfo{author}{\bibfnamefont{S.}~\bibnamefont{Gallego}} \bibnamefont{and}
  \bibinfo{author}{\bibfnamefont{M.}~\bibnamefont{Munoz}},
  \bibinfo{journal}{Surface science} \textbf{\bibinfo{volume}{423}},
  \bibinfo{pages}{324} (\bibinfo{year}{1999}).

\bibitem[{\citenamefont{Chico et~al.}(2004)\citenamefont{Chico, L\'opez-Sancho,
  and Mu\~noz}}]{CLM04}
\bibinfo{author}{\bibfnamefont{L.}~\bibnamefont{Chico}},
  \bibinfo{author}{\bibfnamefont{M.~P.} \bibnamefont{L\'opez-Sancho}},
  \bibnamefont{and} \bibinfo{author}{\bibfnamefont{M.~C.}
  \bibnamefont{Mu\~noz}}, \bibinfo{journal}{Phys. Rev. Lett.}
  \textbf{\bibinfo{volume}{93}}, \bibinfo{pages}{176402}
  (\bibinfo{year}{2004}).

\bibitem[{\citenamefont{Huertas-Hernando
  et~al.}(2006)\citenamefont{Huertas-Hernando, Guinea, and Brataas}}]{HGB06}
\bibinfo{author}{\bibfnamefont{D.}~\bibnamefont{Huertas-Hernando}},
  \bibinfo{author}{\bibfnamefont{F.}~\bibnamefont{Guinea}}, \bibnamefont{and}
  \bibinfo{author}{\bibfnamefont{A.}~\bibnamefont{Brataas}},
  \bibinfo{journal}{Phys. Rev. B} \textbf{\bibinfo{volume}{74}},
  \bibinfo{pages}{155426} (\bibinfo{year}{2006}).

\bibitem[{\citenamefont{Soler et~al.}(2002)\citenamefont{Soler, Artacho, Gale,
  Garc\'ia, Junquera, Ordej\'on, and S\'anchez-Portal}}]{soler}
\bibinfo{author}{\bibfnamefont{J.}~\bibnamefont{Soler}},
  \bibinfo{author}{\bibfnamefont{E.}~\bibnamefont{Artacho}},
  \bibinfo{author}{\bibfnamefont{J.}~\bibnamefont{Gale}},
  \bibinfo{author}{\bibfnamefont{A.}~\bibnamefont{Garc\'ia}},
  \bibinfo{author}{\bibfnamefont{J.}~\bibnamefont{Junquera}},
  \bibinfo{author}{\bibfnamefont{P.}~\bibnamefont{Ordej\'on}},
  \bibnamefont{and}
  \bibinfo{author}{\bibfnamefont{D.}~\bibnamefont{S\'anchez-Portal}},
  \bibinfo{journal}{J. Phys.: Condens.Matter} \textbf{\bibinfo{volume}{14}},
  \bibinfo{pages}{2745} (\bibinfo{year}{2002}).

\bibitem[{\citenamefont{Artacho et~al.}(2008)\citenamefont{Artacho, Anglada,
  Dieguez, Gale, Garc\'ia, Junquera, Martin, Ordej\'on, Pruneda,
  S\'anchez-Portal et~al.}}]{artacho}
\bibinfo{author}{\bibfnamefont{E.}~\bibnamefont{Artacho}},
  \bibinfo{author}{\bibfnamefont{E.}~\bibnamefont{Anglada}},
  \bibinfo{author}{\bibfnamefont{O.}~\bibnamefont{Dieguez}},
  \bibinfo{author}{\bibfnamefont{J.}~\bibnamefont{Gale}},
  \bibinfo{author}{\bibfnamefont{A.}~\bibnamefont{Garc\'ia}},
  \bibinfo{author}{\bibfnamefont{J.}~\bibnamefont{Junquera}},
  \bibinfo{author}{\bibfnamefont{R.}~\bibnamefont{Martin}},
  \bibinfo{author}{\bibfnamefont{P.}~\bibnamefont{Ordej\'on}},
  \bibinfo{author}{\bibfnamefont{J.~M.} \bibnamefont{Pruneda}},
  \bibinfo{author}{\bibfnamefont{D.}~\bibnamefont{S\'anchez-Portal}},
  \bibnamefont{et~al.}, \bibinfo{journal}{J. Phys.: Condens.Matter}
  \textbf{\bibinfo{volume}{20}}, \bibinfo{pages}{064208}
  (\bibinfo{year}{2008}).

\bibitem[{\citenamefont{Fern\'andez-Seivane
  et~al.}(2006)\citenamefont{Fern\'andez-Seivane, Oliveira, Sanvito, and
  Ferrer}}]{seivane}
\bibinfo{author}{\bibfnamefont{L.}~\bibnamefont{Fern\'andez-Seivane}},
  \bibinfo{author}{\bibfnamefont{M.}~\bibnamefont{Oliveira}},
  \bibinfo{author}{\bibfnamefont{S.}~\bibnamefont{Sanvito}}, \bibnamefont{and}
  \bibinfo{author}{\bibfnamefont{J.}~\bibnamefont{Ferrer}},
  \bibinfo{journal}{J. Phys.: Condens.Matter} \textbf{\bibinfo{volume}{18}},
  \bibinfo{pages}{7999} (\bibinfo{year}{2006}).

\bibitem[{\citenamefont{Ceperley and Alder}(1980)}]{ceperly}
\bibinfo{author}{\bibfnamefont{D.}~\bibnamefont{Ceperley}} \bibnamefont{and}
  \bibinfo{author}{\bibfnamefont{B.~J.} \bibnamefont{Alder}},
  \bibinfo{journal}{Phys. Rev. Lett.} \textbf{\bibinfo{volume}{45}},
  \bibinfo{pages}{566} (\bibinfo{year}{1980}).

\bibitem[{\citenamefont{Perdew and Zunger}(1981)}]{perdew}
\bibinfo{author}{\bibfnamefont{P.}~\bibnamefont{Perdew}} \bibnamefont{and}
  \bibinfo{author}{\bibfnamefont{A.}~\bibnamefont{Zunger}},
  \bibinfo{journal}{Phys. Rev. B} \textbf{\bibinfo{volume}{23}},
  \bibinfo{pages}{5048} (\bibinfo{year}{1981}).

\bibitem[{\citenamefont{Artacho et~al.}(1999)\citenamefont{Artacho,
  S\'anchez-Portal, Ordej\'on, Garc\'ia, and Soler}}]{artacho2}
\bibinfo{author}{\bibfnamefont{E.}~\bibnamefont{Artacho}},
  \bibinfo{author}{\bibfnamefont{D.}~\bibnamefont{S\'anchez-Portal}},
  \bibinfo{author}{\bibfnamefont{P.}~\bibnamefont{Ordej\'on}},
  \bibinfo{author}{\bibfnamefont{A.}~\bibnamefont{Garc\'ia}}, \bibnamefont{and}
  \bibinfo{author}{\bibfnamefont{J.}~\bibnamefont{Soler}},
  \bibinfo{journal}{Phys. Status Solidi B} \textbf{\bibinfo{volume}{215}},
  \bibinfo{pages}{809} (\bibinfo{year}{1999}).

\bibitem[{\citenamefont{Bromley et~al.}(1972)\citenamefont{Bromley, Murray, and
  Yoffe}}]{BMY72}
\bibinfo{author}{\bibfnamefont{R.~A.} \bibnamefont{Bromley}},
  \bibinfo{author}{\bibfnamefont{R.~B.} \bibnamefont{Murray}},
  \bibnamefont{and} \bibinfo{author}{\bibfnamefont{A.~D.} \bibnamefont{Yoffe}},
  \bibinfo{journal}{J. Phys. C: Solid State Phys.}
  \textbf{\bibinfo{volume}{5}}, \bibinfo{pages}{759} (\bibinfo{year}{1972}).

\bibitem[{\citenamefont{Schutte et~al.}(1987)\citenamefont{Schutte, Boer, and
  Jellinek}}]{schutte}
\bibinfo{author}{\bibfnamefont{W.}~\bibnamefont{Schutte}},
  \bibinfo{author}{\bibfnamefont{J.}~\bibnamefont{Boer}}, \bibnamefont{and}
  \bibinfo{author}{\bibfnamefont{F.}~\bibnamefont{Jellinek}},
  \bibinfo{journal}{J. Solid State Chem.} \textbf{\bibinfo{volume}{70}},
  \bibinfo{pages}{207} (\bibinfo{year}{1987}).

\bibitem[{foo()}]{footnote}
\bibinfo{note}{We notice that the TB parameters used in this work lead to a
  trigonal warping of the conduction band which is rotated $\pi/3$ with respect
  to the DFT bands. This fact does not affect the results discussed here.}

\bibitem[{\citenamefont{{Alidoust} et~al.}(2013)\citenamefont{{Alidoust},
  {Bian}, {Xu}, {Sankar}, {Neupane}, {Liu}, {Belopolski}, {Qu}, {Denlinger},
  {Chou} et~al.}}]{AH13}
\bibinfo{author}{\bibfnamefont{N.}~\bibnamefont{{Alidoust}}},
  \bibinfo{author}{\bibfnamefont{G.}~\bibnamefont{{Bian}}},
  \bibinfo{author}{\bibfnamefont{S.-Y.} \bibnamefont{{Xu}}},
  \bibinfo{author}{\bibfnamefont{R.}~\bibnamefont{{Sankar}}},
  \bibinfo{author}{\bibfnamefont{M.}~\bibnamefont{{Neupane}}},
  \bibinfo{author}{\bibfnamefont{C.}~\bibnamefont{{Liu}}},
  \bibinfo{author}{\bibfnamefont{I.}~\bibnamefont{{Belopolski}}},
  \bibinfo{author}{\bibfnamefont{D.-X.} \bibnamefont{{Qu}}},
  \bibinfo{author}{\bibfnamefont{J.~D.} \bibnamefont{{Denlinger}}},
  \bibinfo{author}{\bibfnamefont{F.-C.} \bibnamefont{{Chou}}},
  \bibnamefont{et~al.}, \bibinfo{journal}{ArXiv e-prints}
  (\bibinfo{year}{2013}), \eprint{1312.7631}.

\bibitem[{\citenamefont{{Yuan} et~al.}(2014)\citenamefont{{Yuan}, {Wang},
  {Lian}, {Zhang}, {Fang}, {Shen}, {Xu}, {Xu}, {Zhang}, {Hwang} et~al.}}]{YC14}
\bibinfo{author}{\bibfnamefont{H.}~\bibnamefont{{Yuan}}},
  \bibinfo{author}{\bibfnamefont{X.}~\bibnamefont{{Wang}}},
  \bibinfo{author}{\bibfnamefont{B.}~\bibnamefont{{Lian}}},
  \bibinfo{author}{\bibfnamefont{H.}~\bibnamefont{{Zhang}}},
  \bibinfo{author}{\bibfnamefont{X.}~\bibnamefont{{Fang}}},
  \bibinfo{author}{\bibfnamefont{B.}~\bibnamefont{{Shen}}},
  \bibinfo{author}{\bibfnamefont{G.}~\bibnamefont{{Xu}}},
  \bibinfo{author}{\bibfnamefont{Y.}~\bibnamefont{{Xu}}},
  \bibinfo{author}{\bibfnamefont{S.-C.} \bibnamefont{{Zhang}}},
  \bibinfo{author}{\bibfnamefont{H.~Y.} \bibnamefont{{Hwang}}},
  \bibnamefont{et~al.}, \bibinfo{journal}{ArXiv e-prints}
  (\bibinfo{year}{2014}), \eprint{1403.2696}.

\bibitem[{\citenamefont{Castellanos-Gomez
  et~al.}(2013)\citenamefont{Castellanos-Gomez, Rold\'an, Cappelluti, Buscema,
  Guinea, van~der Zant, and Steele}}]{CS13}
\bibinfo{author}{\bibfnamefont{A.}~\bibnamefont{Castellanos-Gomez}},
  \bibinfo{author}{\bibfnamefont{R.}~\bibnamefont{Rold\'an}},
  \bibinfo{author}{\bibfnamefont{E.}~\bibnamefont{Cappelluti}},
  \bibinfo{author}{\bibfnamefont{M.}~\bibnamefont{Buscema}},
  \bibinfo{author}{\bibfnamefont{F.}~\bibnamefont{Guinea}},
  \bibinfo{author}{\bibfnamefont{H.~S.~J.} \bibnamefont{van~der Zant}},
  \bibnamefont{and} \bibinfo{author}{\bibfnamefont{G.~A.}
  \bibnamefont{Steele}}, \bibinfo{journal}{Nano Letters}
  \textbf{\bibinfo{volume}{13}}, \bibinfo{pages}{5361} (\bibinfo{year}{2013}).

\bibitem[{\citenamefont{Kleinman}(1980)}]{kleinman_1980}
\bibinfo{author}{\bibfnamefont{L.}~\bibnamefont{Kleinman}},
  \bibinfo{journal}{Physical Review B} \textbf{\bibinfo{volume}{21}}
  (\bibinfo{year}{1980}).

\bibitem[{\citenamefont{Gong et~al.}(2013)\citenamefont{Gong, Liu, Yu, Xiao,
  Cui, Xu, and Yao}}]{GY13}
\bibinfo{author}{\bibfnamefont{Z.}~\bibnamefont{Gong}},
  \bibinfo{author}{\bibfnamefont{G.-B.} \bibnamefont{Liu}},
  \bibinfo{author}{\bibfnamefont{H.}~\bibnamefont{Yu}},
  \bibinfo{author}{\bibfnamefont{D.}~\bibnamefont{Xiao}},
  \bibinfo{author}{\bibfnamefont{X.}~\bibnamefont{Cui}},
  \bibinfo{author}{\bibfnamefont{X.}~\bibnamefont{Xu}}, \bibnamefont{and}
  \bibinfo{author}{\bibfnamefont{W.}~\bibnamefont{Yao}},
  \bibinfo{journal}{Nature Communications} \textbf{\bibinfo{volume}{4}}
  (\bibinfo{year}{2013}).

\end{thebibliography}

\end{document}